\documentclass[12pt,twoside]{article}
\usepackage{fleqn,espcrc1,psfig,epsf}

\newcommand{\vb}{{\vec b}}
\newcommand{\vp}{{\vec p}}

\newcommand{\vx}{{\vec x}}

\newcommand{\beq}{\begin{equation}}
\newcommand{\eeq}[1]{\label{#1} \end{equation}}

\newcommand{\lton}{\mathrel{\lower.9ex
                  \hbox{$\stackrel{\displaystyle <}{\sim}$}}}
\newcommand{\ee}{\end{equation}} \newcommand{\ben}{\begin{enumerate}}
\newcommand{\een}{\end{enumerate}} \newcommand{\bit}{\begin{itemize}}
\newcommand{\eit}{\end{itemize}} \newcommand{\bc}{\begin{center}}
\newcommand{\ec}{\end{center}} \newcommand{\bea}{\begin{eqnarray}}
\newcommand{\eea}{\end{eqnarray}}
\newcommand{\beqar}{\begin{eqnarray}}
\newcommand{\eeqar}[1]{\label{#1} \end{eqnarray}}

\title{
The Next Yukawa Phase of QCD at RHIC
}

\author{Miklos Gyulassy  
        \\[1ex]
Physics Department, Columbia University,
New York, NY 10027}
\begin{document}
\maketitle
\begin{abstract}
QCD predicts the existence of a new partonic Yukawa phase 
above a critical temperature $T_c\sim
150$ MeV. 
I review some of the key observables which will be
used in heavy ion reactions 
to  search for this  new  phase at RHIC. 
Systematics of  collective observables ($N_{ch},E_\perp)$, flow patterns,
meson interferometry, jet quenching, and $J/\psi$
suppression are discussed. This talk is updated with first data from RHIC.

\end{abstract}


\section{Yukawa in QCD}
In 1935 H. Yukawa\cite{Yuk35} proposed a theory of nuclear forces based
on the  exchange of a massive boson
\beq
V_Y(r,m)= \alpha_{eff} \frac{e^{-mr}}{r}
\eeq{yukawa}
He estimated that $m \sim 100$ MeV to account 
for the short range $\sim 2$ fm of the nuclear force, and in 
1947 Powell  confirmed his  theory.
Yukawa's meson 
theory forms the basis for the current
effective theory of nuclear forces\cite{Bonn}.

In the field of electrolytes, Debye and H\"uckle\cite{Debye}
had already discovered the Yukawa potential in a completely
different  context.
The  polarizability of the medium in the presence of an external charge
density leads to a non-linear  self consistent equation
with the same solution as   eq. (\ref{yukawa}), but in that case
 the effective mass  is  the Debye electric screening mass
\beq
\mu^2=4\pi \sum_q q^2 |n_q|/T
\; \; . \eeq{mudeb}
A conductive medium transforms Coulomb power law forces into the
Yukawa form.

In nuclear theory, the Yukawa meson mass results from the finite gap
of the elementary excitations (pions, ...)  of the physical QCD
vacuum. In this talk, I discuss current efforts to try to drill a
perturbative hole into the nonperturbative vacuum using RHIC to see
the breakdown of Yukawa's meson theory.  As we review below, in the
deconfined, chirally symmetric phase of QCD at high temperatures, the
Debye-Huckle mechanism transforms the mesonic Yukawa potentials into a
color-electric screened gluonic Yukawa potential in a quark-gluon
plasma (QGP).

\begin{figure}
\vspace{-0.4in}
\hspace{0.1in}\epsfxsize= 3.8 in  
\epsfysize= 3.9 in
\centerline{\epsfbox{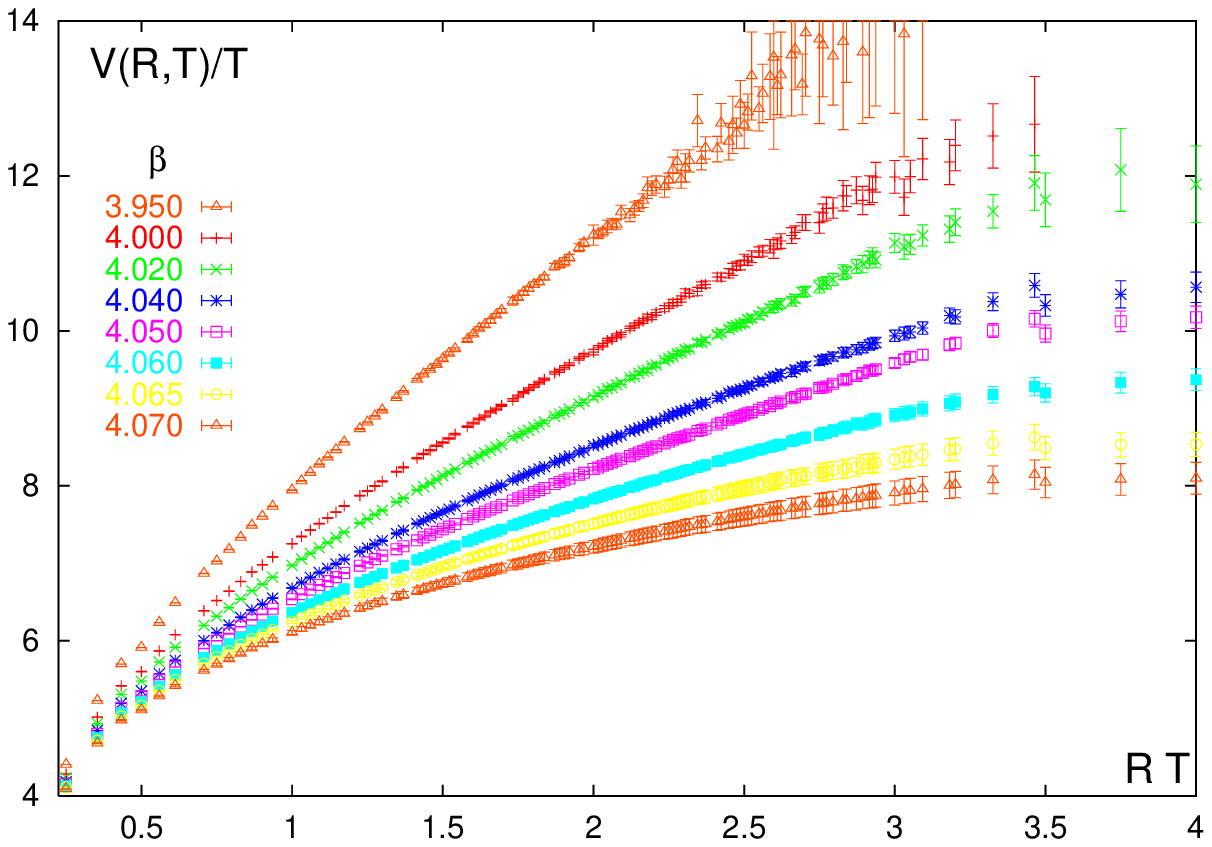}}
\vspace{-0.45in}
\centerline{\epsfxsize= 4 in  
\epsfysize= 3.9 in\epsfbox{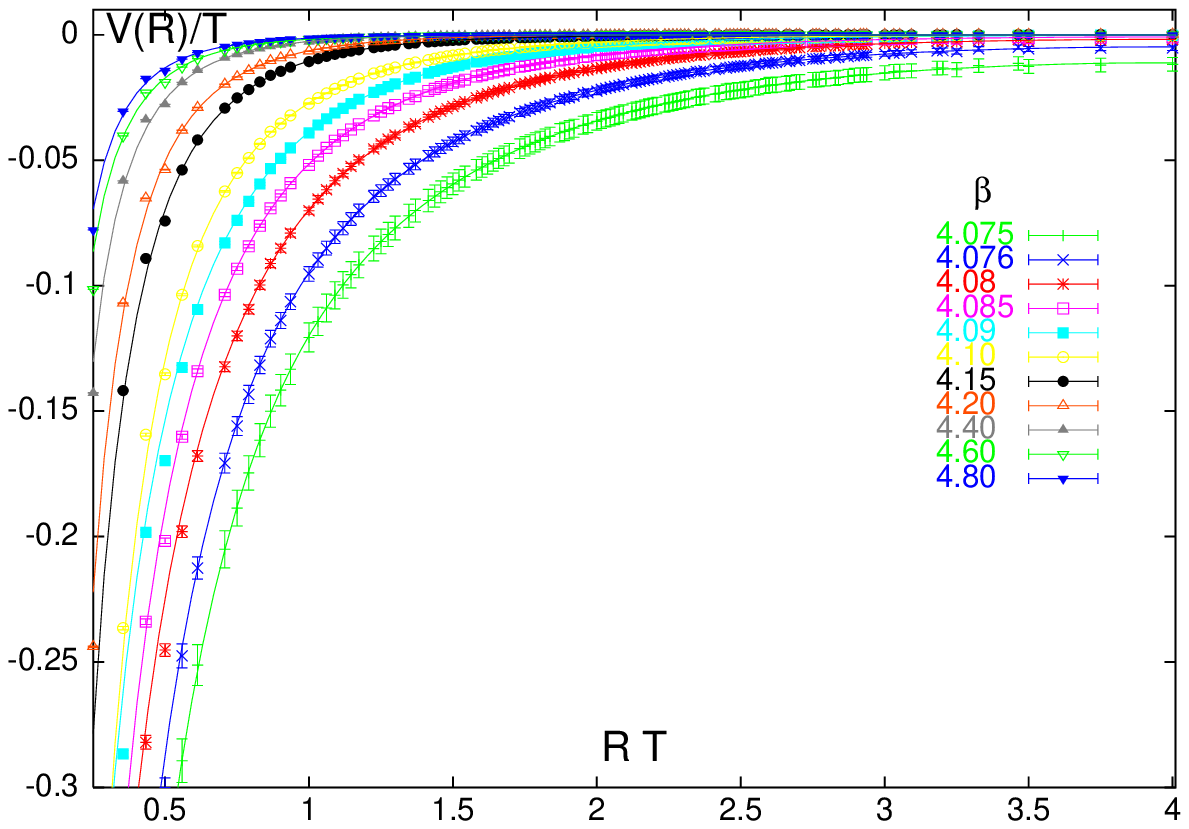}}
\vspace{-0.3in}
\caption{ The heavy quark potential in the confined 
phase of SU(2) quenched
QCD (top)  compared to that in the deconfined Yukawa phase
above \protect{$T_c$} (bottom). Results are for a $32^3\times 4$ lattice
from Karsch et al\protect{\cite{Kaczmarek:1999mm}}.
Eq.(\protect{\ref{Gao}}) fits the confined lattice "data" well,
but the QCD string tension decreases more rapidly near $T_c$.
For $T>T_c$ the potential is  screened by the deconfined
gluons (in this quenched calculation) and  acquires
 the generalized Yukawa form
(\protect{\ref{free_d}}). Here $T\approx T_c \exp(11/6(\beta-\beta_c))$ 
increases with $\beta$ and  $\beta_c=4.0729$.
}
\label{fig:confine}
\end{figure}

In QCD, the color potential between partons is approximately
Coulombic at small distances due to the 
asymptotic freedom property of non-Abelian gauge theories.
 However, below a critical temperature, $T_c\sim 150$ MeV,
the effective potential between the colored partons  has
a long range linear confining term
$\kappa r$ with a huge  "string'' tension, $\kappa\sim 1$ GeV/fm.
In this confining phase of QCD, the heavy $q\bar{q}$ potential 
is well parameterized by  the
L\"uscher form\cite{Lusch}
\beq
V_{L}(r,0) = -\frac{\alpha_L}{r}+\kappa r
\;\; ,\eeq{lusch}
(as long as dynamical quark pair production is ignored).
The Coulombic part, with strength $\alpha_L=\pi/12$,  arises 
from the zero point quantum fluctuations of the string. 
As the temperature increases, but remains below
the deconfinement transition, $T<T_c$, the enhanced 
fluctuations due to thermal agitation of the string
modifies the effective potential into the 
approximate Gao form
\cite{Gao}
\begin{eqnarray}
V_G(r,T) & = & -\frac{\alpha_L}{r} \left[ 1 - \frac{2}{\pi} \tan^{-1}
(2rT)\right]
 +\left[ \kappa - \frac{\pi}{3} T^2(1 - \frac{2}{\pi}
\tan^{-1}(\frac{1}{2rT})\right] r +\cdots
\label{Gao} 
\end{eqnarray}
The decrease of 
the effective string tension, $\kappa(T)$, predicted above
has been measured via
lattice QCD calculations\cite{Kaczmarek:1999mm}  as
shown in Fig.(\ref{fig:confine}). However, the ``measured''
 string tension is found to
decreases faster than predicted in eq.(\ref{Gao}) near
the  critical temperature. Note that in Fig.(\ref{fig:confine})
$T\approx T_c \exp(11/6(\beta-\beta_c))$ 
with $\beta_c=4.0729$ for this lattice.

\begin{figure}
\vspace{-0.7in}
\epsfxsize= 3 in  
\epsfysize= 4 in
\centerline{\epsfbox{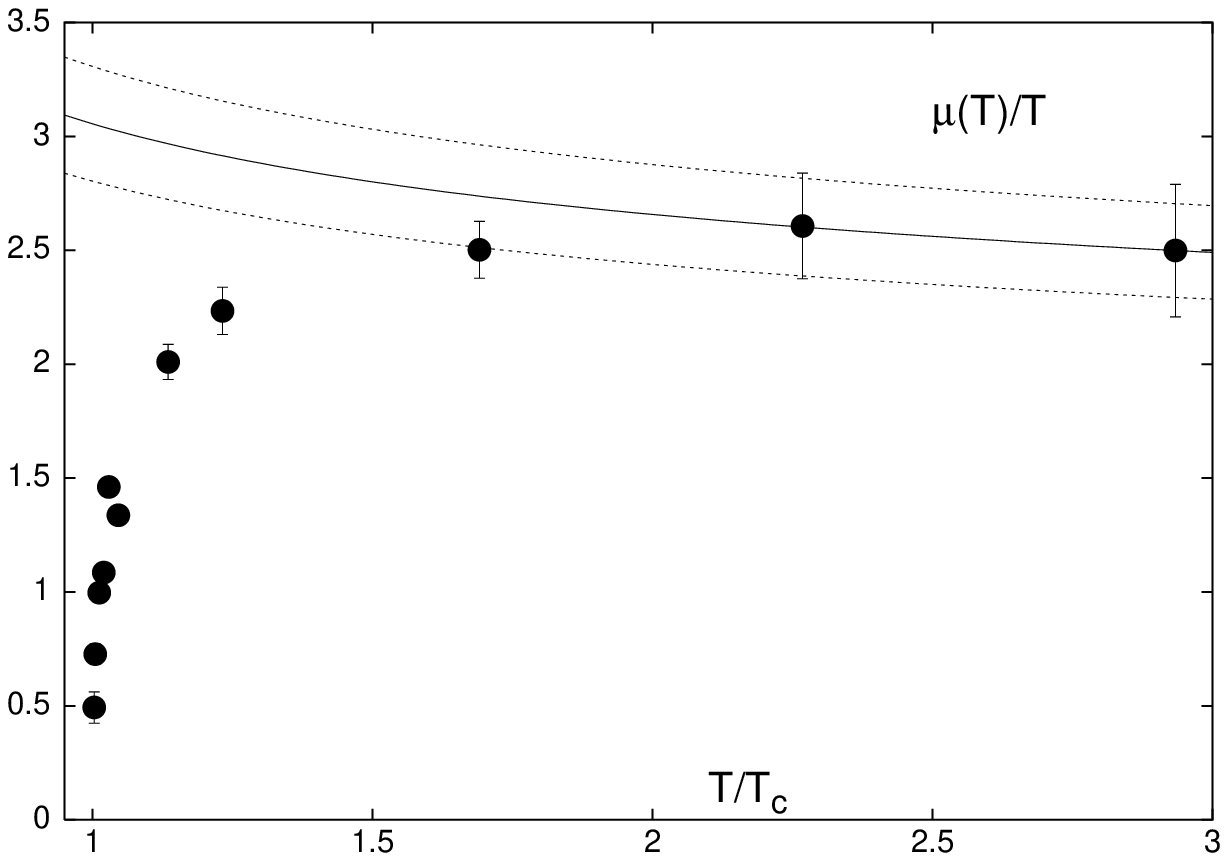}\epsfxsize= 3 in \epsfysize= 4 in
  \epsfbox{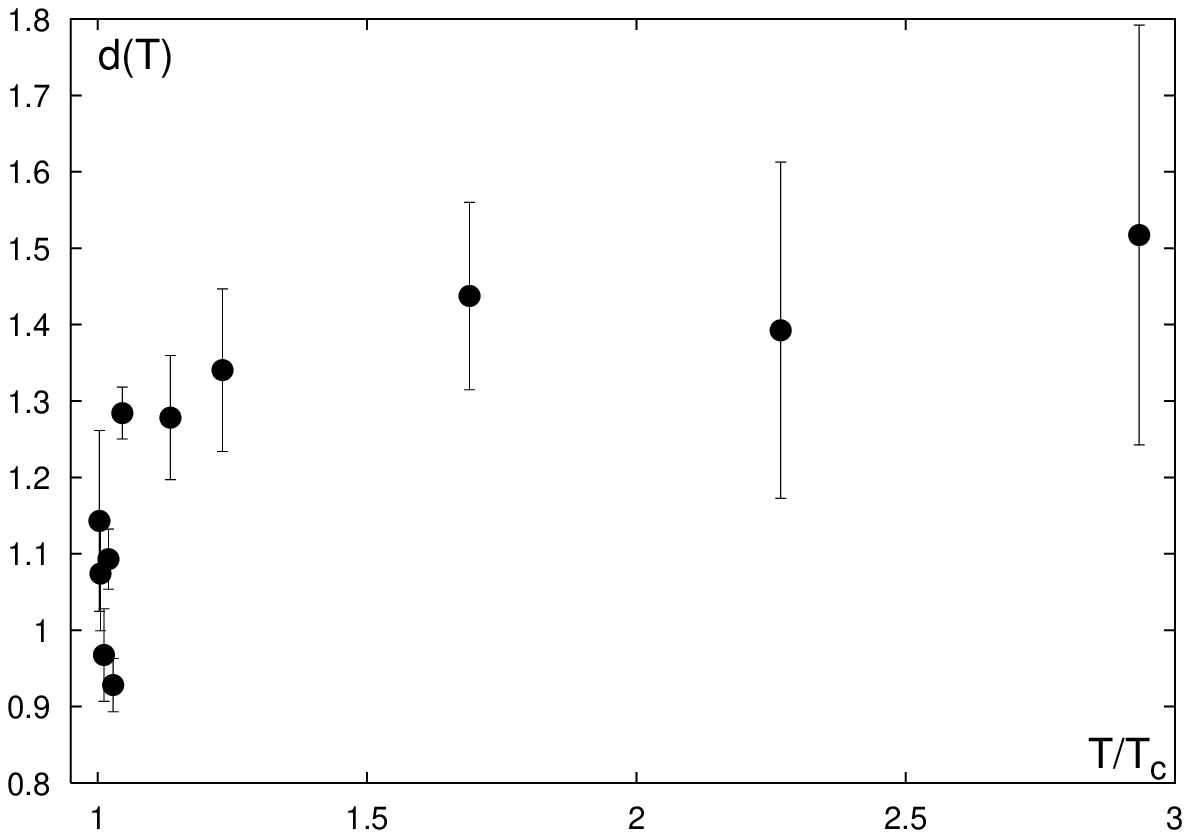}}
\vspace{-0.4in}
\caption{The chromo-electric Debye screening mass, $\mu(T)=2 m_E(T)$   and the
effective exponent, $d(T)$, of the effective Yukawa potential 
in the deconfined  phase versus $T/T_c$ is shown from 
Karsch et al \protect{\cite{Kaczmarek:1999mm}} for SU(2) quenched
QCD. The lines show expected dependence based on thermal  pQCD.
Note that near $T_c$, the range $2/\mu$ is much larger than 
predicted by pQCD and that $d\approx 1$ implies an especially
long range interaction there.
}
\label{fig:mu}
\end{figure}

For temperatures above $T_c$, the lattice data in  Fig.\ref{fig:confine}
reveal the predicted   Yukawa phase of QCD is predicted.
The  heavy quark potential mutates into a short range
generalized Yukawa form, which on the lattice is fit with the form 
\beq
V_L(r,T,d) = - \frac{\alpha(T) T}{(rT)^{d_L/2}} e^{-\mu_L(T)r/2}
\;\; .\label{free_d}
\eeq{vy}
Note that $d_L=2, \mu_L(T)=2 m_E(T)$ correspond to the ideal Yukawa
form. The perturbative
thermal QCD 
 chromo-electric  Debye mass $m_E=\mu(T)/2$ is \cite{Rebhan:1994mx} 
\beq
m_E(T) = g(T)T \left(\frac{N_c}{3}
+\frac{N_F}{6}\right)^{1/2}
\eeq{pertmu}
  for  $N_c$ colors and  $N_F$ flavors. For $N_c=2,N_f=0$ in Fig.
(\ref{fig:confine}), we expect $\mu_L(T)= 2m_E=1.6 g(T) T$
as shown by the solid line in Fig.(\ref{fig:mu}).
The fits to the lattice QCD data in  Fig.(\ref{fig:confine} b)
from \cite{Kaczmarek:1999mm} show that for $T>2 T_c$
 $\mu_L\approx 2.5 T$ is not far from the pQCD estimate.
However, the exponent $d_L\approx 1.5$ is significantly below
the value 2 expected from pQCD. An even more striking nonperturbative
deviation is seen near $T_c$, at which point $d<1$
and $\mu\sim T/2$. This suggests a rather long range interaction
that may be the precursor of the confinement transition.

\begin{figure}
\vspace{-0.7in}
\epsfxsize= 3. in  
\centerline{\hspace{0.15in}\epsfbox{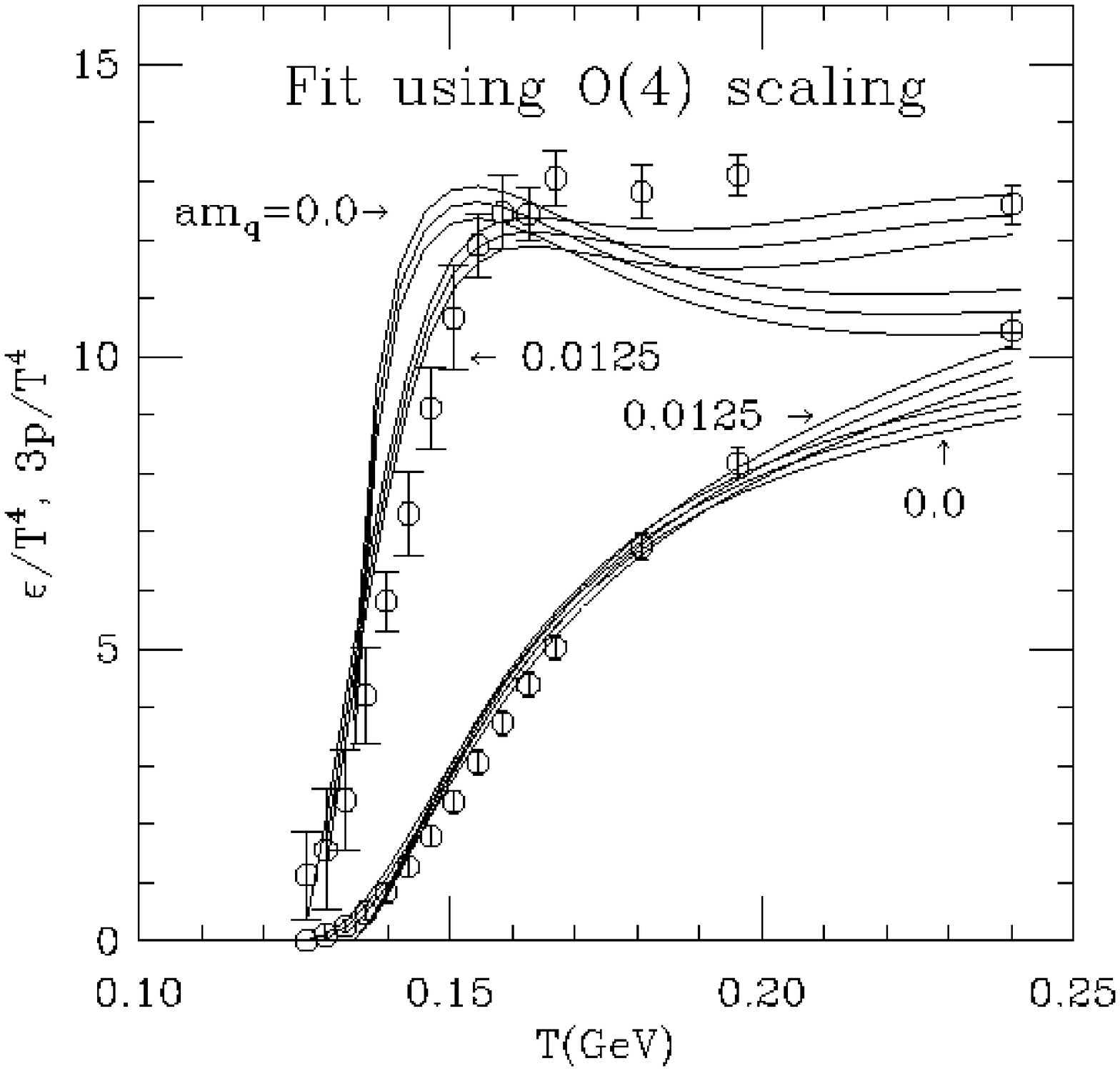}\epsfxsize= 3. in   
\hspace{-0.15in} \epsfbox{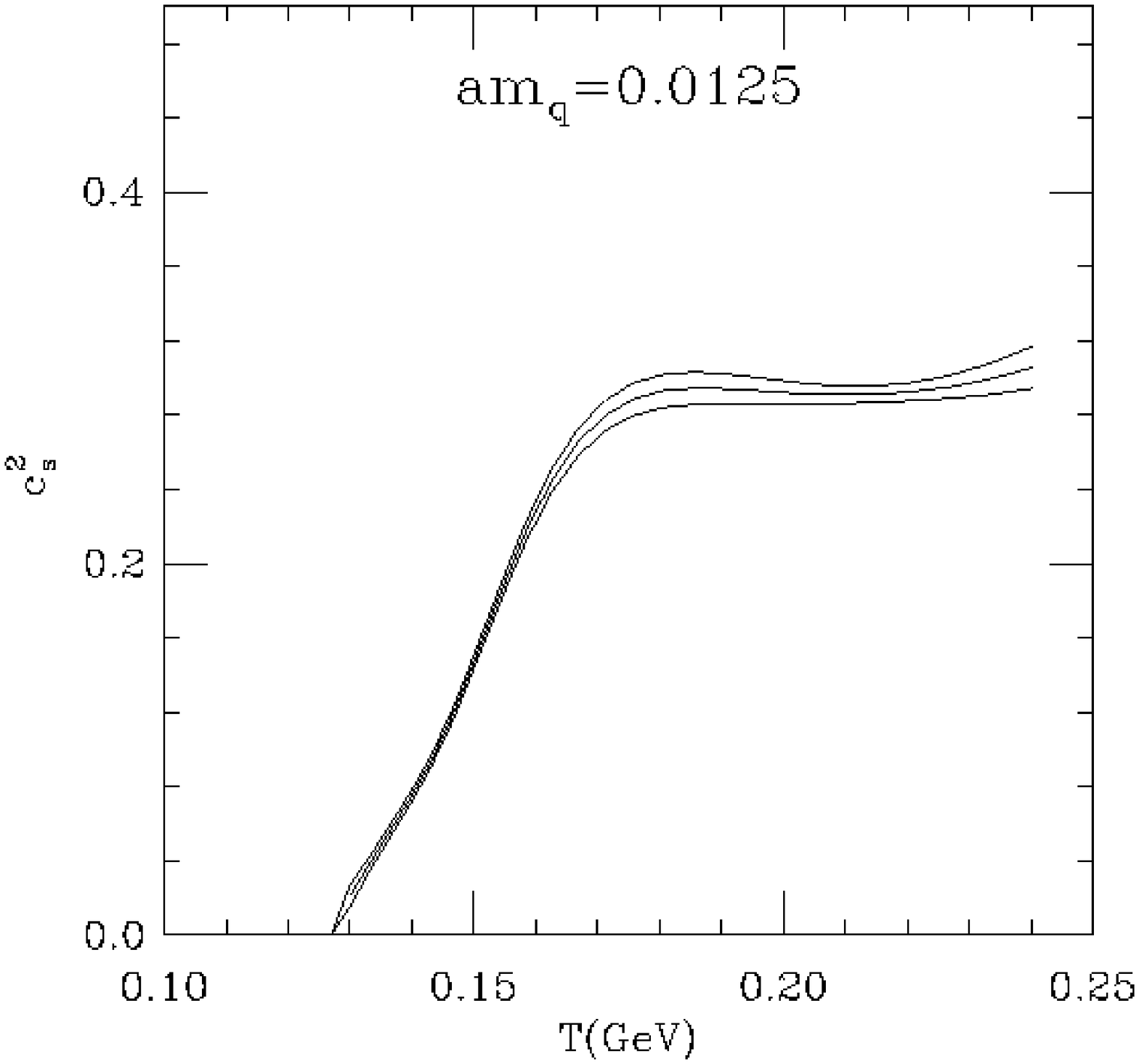}}
\vspace{-0.3in}
\caption{Thermodynamic 
energy density ($\epsilon/T^4$ top curves left), pressure
($3p/T^4$ lower curves left) and speed of sound squared
(right) from lattice QCD
(2 flavor $12^3\times 6$) from the MILC collab \protect{\cite{Bernard:1997cs}}.
The curves are zero quark mass extrapolations.
Note the rapid reduction of the pressure and speed of sound
as $T_c$ is approached from above.}
\label{fig:eos}
\end{figure}

The predicted thermodynamic properties of the QCD Yukawa  phase
are shown in Fig.\ref{fig:eos} from ref.\cite{Bernard:1997cs}
for 2 flavor $12^3\times 6$ lQCD.
A present caveat about  all lattice results
is that the pion is still too massive to allow  contact
with the ``known'' thermodynamic
 properties of ordinary hadronic/nuclear matter below $T_c$.
Two striking features of Fig.(\ref{fig:eos}) suggest two key
observable signatures of this phase transition in nuclear collisions.
First, the entropy density $\sigma(T)=(\epsilon+p)/T$ increase
very rapidly with $T$ in a narrow interval $\Delta T/T_c< 0.1$.
Second, the plasma becomes extremely soft $p/\epsilon\ll 1$ and
$c_s^2\ll1$ near $T_c$. As we review below, the first feature may lead to
time delay (the QGP stall) 
measurable via hadronic interferometry. The second
feature may lead to interesting non-linear collective 
flow observables in nuclear collisions.
The experimental verification of these fundamental 
predictions of QCD is one of the primary goals of 
the heavy ion experimental program at Brookhaven and CERN.
In the following sections, I review first  how deep into the QGP
phase RHIC may be able to reach,
and then discuss several signatures that will  be used to test
the QCD  predictions in such experiments.

\section{Initial conditions in A+A}

In order to see the partonic Yukawa phase, we must first
heat the vacuum to about 100 times  the energy density of nuclear matter.
At SPS energies the low $p_\perp$ physics makes it impossible
to predict  the initial conditions.
However, with increasing energy pQCD begins to provide increasingly 
more reliable theoretical basis for predicting those initial conditions.
For highly boosted nuclei with  $E_{cm}> 100
m_N$, time dilation effectively freezes out the  quantum chromo fluctuations 
inside the nuclei while the two pass through each other. 
Au beams at collider  energies can be thought of as 
 well collimated, ultra dense
beams of partons. This  (
chromo Weizsacker-Williams\cite{McLerran:1994ni})  gluon 
cloud contains a very large number, $G_A(x,p_0^2)\sim A/x^{1+\delta}$,
of almost on-shell collinear gluons with longitudinal momentum fraction
$x=p_0/E_{cm}\ll 1$. As the clouds pass through each other,
many of the (virtual) partons scatter and form a dominantly 
gluonic plasma on a very fast time scale $1/p_\perp \ll 1$ fm/c.
  The number of gluons pairs (mini-jets) extracted from the nuclei
by this mechanism at
 rapidities $y_i$ and transverse momentum $\pm k_\perp$ 
can be calculated in pQCD from the well known expression
\cite{Eskola:2000fc,Eskola:1989yh,Blaizot:1987nc,Gyulassy:1997vt}
\beq
\frac{dN_{AB\rightarrow ggX}}{dy_1dy_2dk_\perp^2}
=K x_1 G_A(x_1,k_\perp^2) x_2 G_B(x_2,k_\perp^2)
\frac{d\sigma_{gg\rightarrow gg}}{dk_\perp^2}
T_{AB}(\vb)
\;\; ,\eeq{mini}
where $x_1=x_\perp(\exp(y_1)+\exp(y_2))$ and
$x_2=x_\perp(\exp(-y_1)+\exp(-y_2))$, with $x_\perp=k_\perp/\surd s$, and where
the pQCD $gg\rightarrow gg $ cross section for scattering with
$t=-k_\perp^2(1+\exp(y_2-y_1))$ and $y_2-y_1=y$ is given by \beqar
\frac{d\sigma^{gg}}{dt}&=& 
\frac{9}{8}
\frac{4\pi\alpha^2}{k_\perp^4}\frac{(1+e^y+e^{-y})^3}{(e^{y/2}+e^{-y/2})^6}
\; \; . \eeqar{ggqcd}
For atomic numbers, $A\gg 1$, the
 geometrical
amplification, $T_{AB}(\vb)\stackrel{<}{\sim} 30/$mb, enhances
by orders of magnitude the gluon density relative to $pp$.
The factor $K\sim 2$ approximates next to leading order contributions.

For symmetric systems, $A+A$, with $G_A\approx A G$,
the inclusive gluon jet production cross section is obtained by integrating
over $y_2$ with $y_1=y$ and $k_\perp$ fixed. 
To about $50\%$ accuracy, the single inclusive gluon rapidity density
in central collisions is approximately\cite{Gyulassy:1997vt}
\beq
        {{dN} \over {dydt}} \approx  {{A^2 \over {\pi R^2}}
        2 N_g(x_\perp,t)}x_\perp  G(x_\perp,t) 
{{d\sigma^{gg}} \over {dt}}\propto A^{4/3}
\eeq{dnint}
where $N_g(x_\perp,t)=\int_{x_\perp}^1 dx G(x,t)$.
This copious mini-jet production mechanism 
is believed to the dominant source of 
 gluon plasma production at RHIC and higher energies.

\begin{figure}
\vspace{1.0in}
\centerline{\epsfxsize= 3.0 in 
\epsfbox{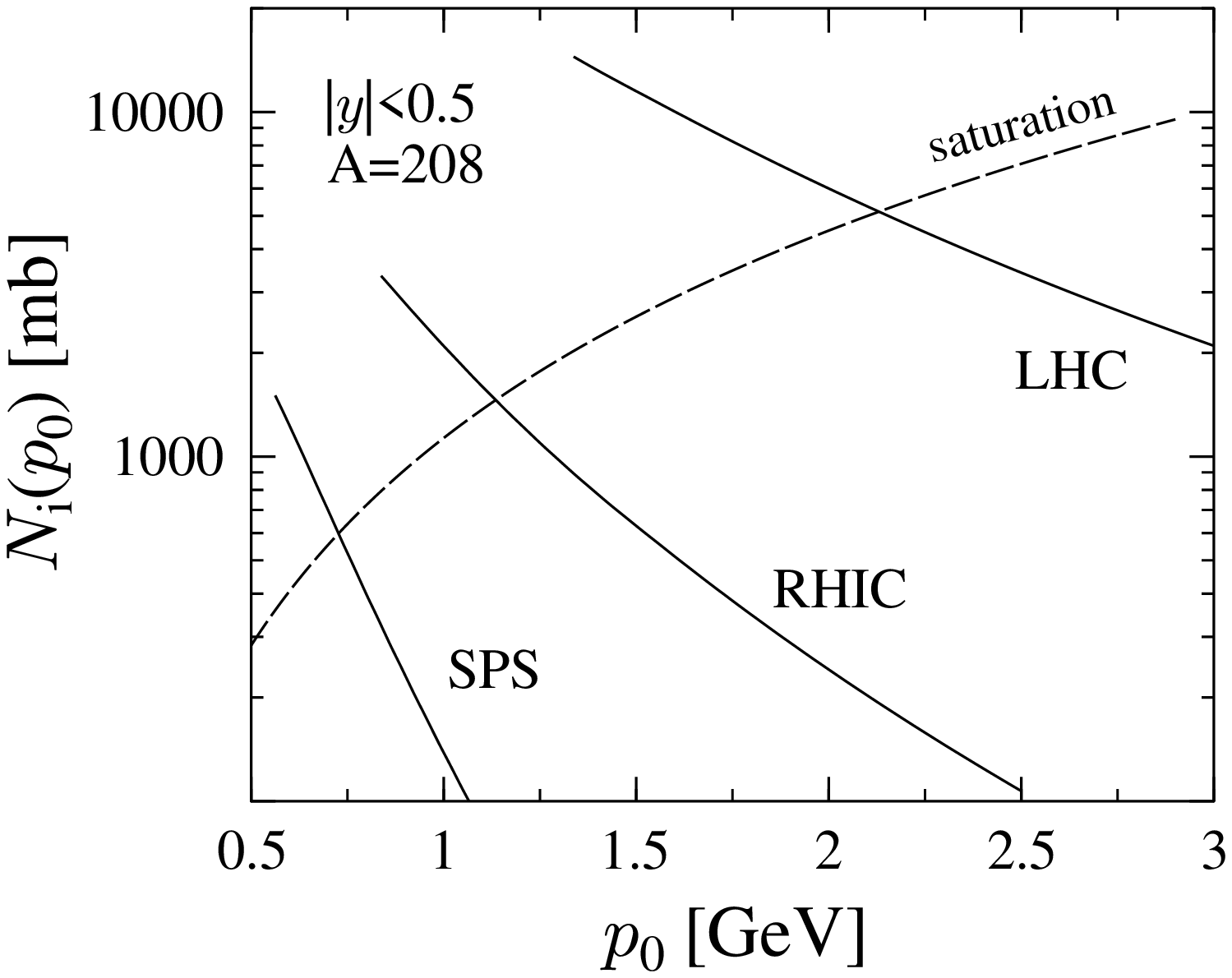}\epsfxsize= 3.0 in 
  \epsfbox{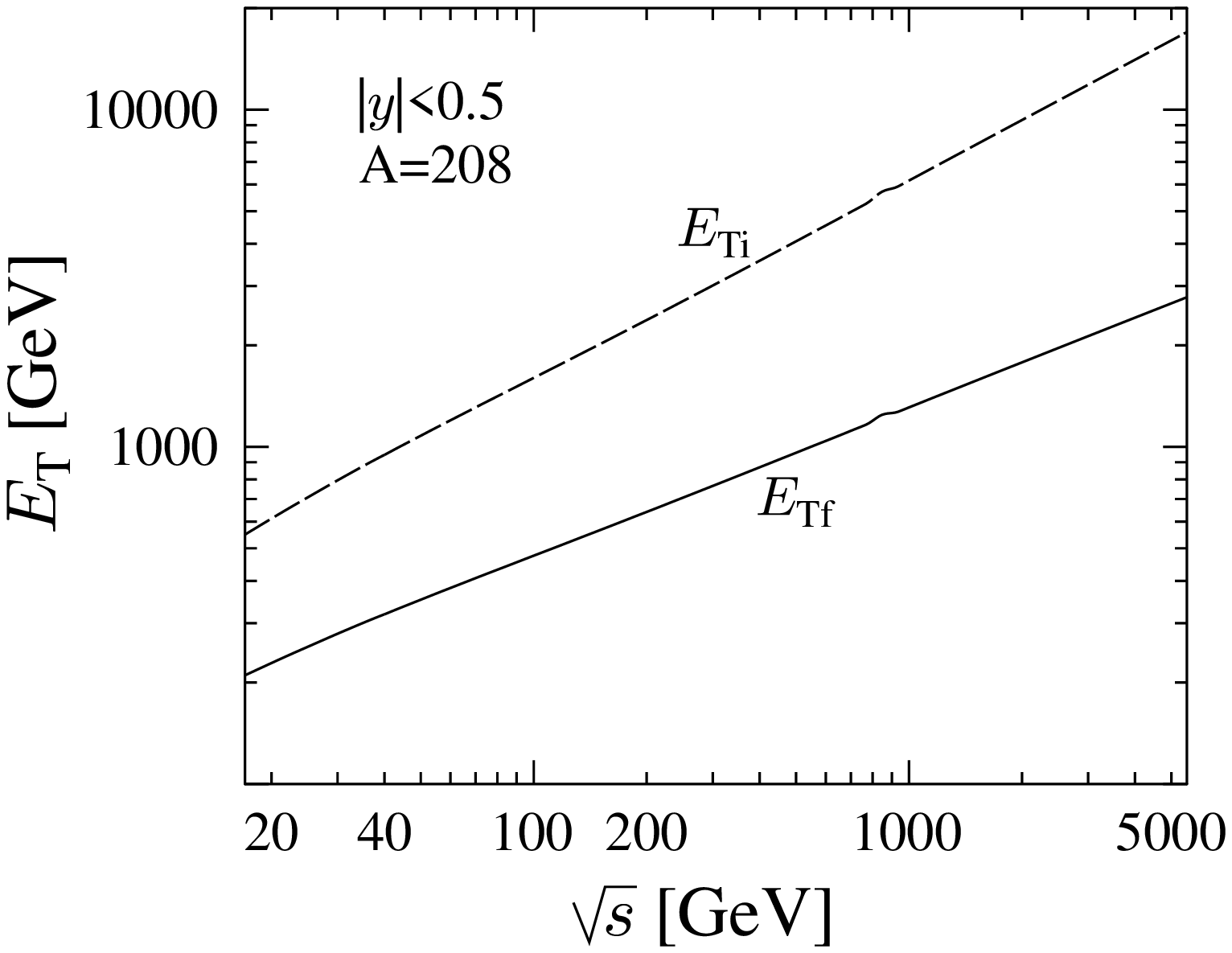}}
\vspace{-2.5in}
\caption{Mini-jet initial rapidity density of gluons, $N_i$,
produced in central A+A collisions as a function of the pQCD cutoff, $p_0$,
 from \protect{\cite{Eskola:2000fc}}. The dashed ``saturation'' curve
corresponds to $(p_0^2R^2)$.
($N_i$ is dimensionless; $[mb]$ is a typo).
 The magnitude of possible
hydrodynamic transverse energy loss due to longitudinal work in an
ideal $p=\epsilon/3$ quark gluon plasma is shown by the
difference between $E_{Ti} $ and    $E_{Tf}$ in the right panel.}
\label{fig:mini}
\end{figure}

Recent upper bound  estimates 
of the total gluon rapidity density as a function of
the CM energy 
from EKRT\cite{Eskola:2000fc}
are shown in Fig.(\ref{fig:mini}). 
The differential yields are integrated down to a transverse momentum scale
$p_0\sim 1-2$ GeV. This scale separates the ``soft"
 nonperturbative beam jet fragmentation domain from
the calculable perturbative one above. 
The curve marked saturation\cite{Blaizot:1987nc}
is an upper bound marking the point where the transverse
gluon density of mini-jets becomes so high that the
newly liberated gluons completely fill the nuclear area,
i.e., $ dN/dy \approx p_0^2 R^2$. A current hypothesis
is that
at that point, higher order
gluon absorption may limit the further increase of
the gluon number. At RHIC energies EKRT
estimates yield up  1500 gluons per unit rapidity.
Our  conservative estimates together with
X.N. Wang\cite{Wang:1991ht,Wang:1997yf} only gives $< 500$  gluons
per unit rapidity when initial and final state radiation is also taken into
account. This lower number is obtained with a fixed $p_0=2$ GeV,
that we  found to be necessary
in order to reproduce pp, $p\bar{p}$ and lower energy
$BA$ data using the HIJING event generator\cite{Wang:1991ht}.
The initial energy density reached in such collisions can be
estimated using the Bjorken formula 
\beq \epsilon(\tau_0) \approx
\frac{1}{\pi R^2\tau_0} \frac{dE_T}{dy} \eeq{bj} For $p_0\sim 1-2$ GeV,
$dE_T/dy\sim 400-2000$ GeV, and therefore $\epsilon(\tau_0 \sim 0.5\;{\rm fm/c}) > 10$ GeV/fm$^3$ 
should be easily reached at RHIC,  well inside  the
the deconfinement phase of QCD. 
 At SPS energies, on the other hand, our  estimates indicate that
nuclear collisions may just reach the 
transition region and cool below it  in  a very short time.

Recently CERN issued a press release\cite{cernhype} 
claiming that ``
We now have  evidence of a new state of matter where quarks and gluons are not confined.''
As discussed below, I disagree with the claim  that
deconfinement has been observed.
What I find compelling is that {\em some} form of matter,
much denser than ever studied before, was created. 
 Inferences about the role of quark and
gluon degrees of freedom cannot be drawn because at
the relatively low momentum scales accessible
at SPS energies, the quarks and gluon degrees of freedom in the dense matter
are mostly not resolvable. 
Even at the highest transverse momentum measured,
 the pion
spectra were shown to be very sensitive to nonperturbative model assumptions
regarding the role of intrinsic momenta
and soft initial state interactions\cite{Gyulassy:1998nc}. The
dynamics of the  non-perturbative beam jet fragmentation and hadronic final
state interactions simply cannot be disentangled at SPS 
energies\cite{Bass:1999vz}. 
While there are many
interesting signatures showing that dense matter was formed at the SPS
(through the non-linear in dependence of several observables on multiplicity or
$A$), the bottom line is that those data have said nothing about whether the
QCD predictions in Figs 1-3 are correct or not.  We need higher
resolution, i.e. energy,  to see the quarks and gluons in action in the new
Yukawa phase. 

\section{First RHIC Data on Multiplicity}
\begin{figure}
\centerline{\epsfxsize= 3.0 in 
\epsfbox{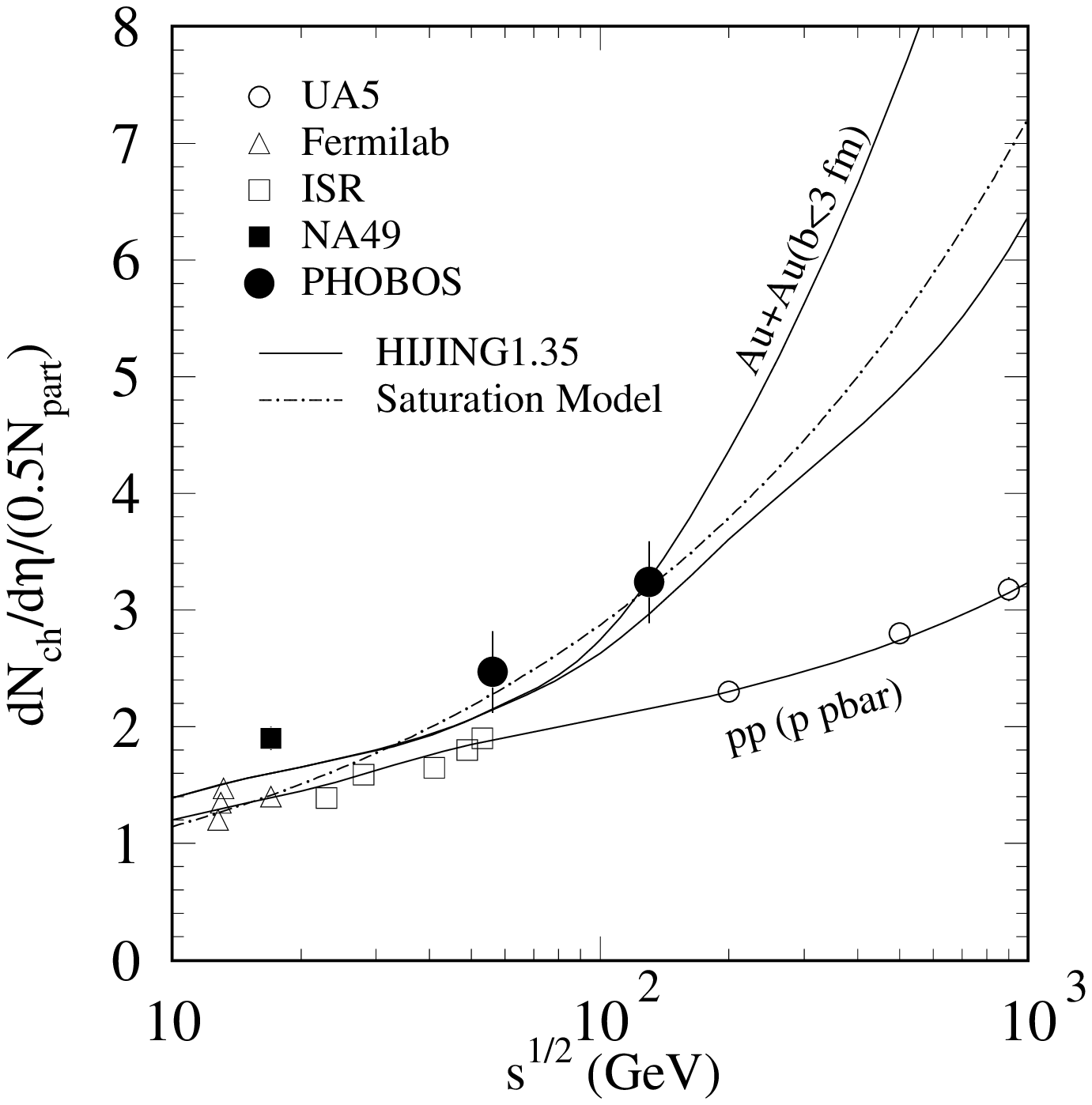}\epsfxsize= 3.0 in 
  \epsfbox{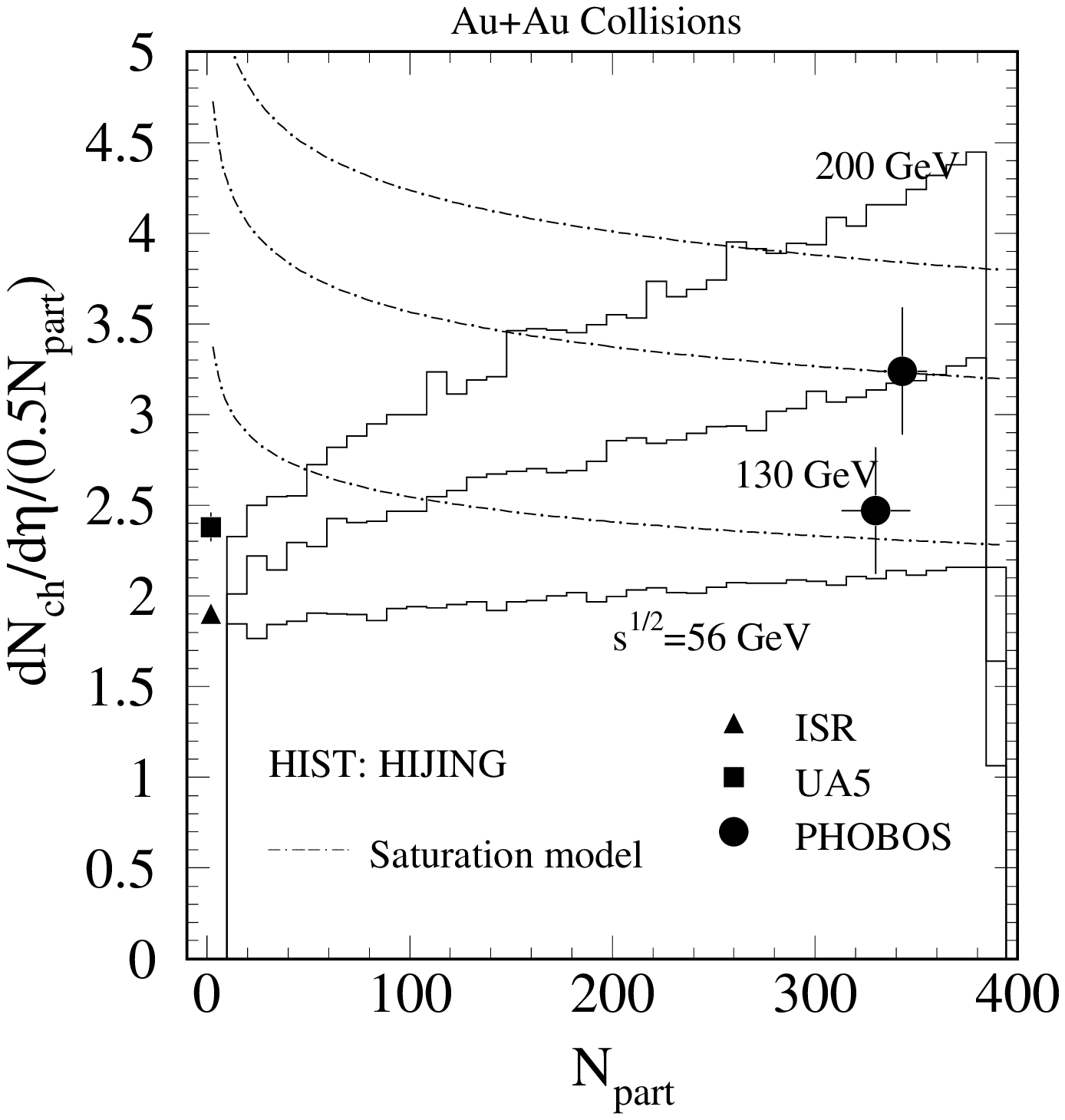}}
\vspace{0.0in}
\caption{Charged particle rapidity density {\em per  participating
baryon pair} from \cite{wg00} 
is shown versus the cm energy per baryon. The  PHOBOS
data\protect\cite{phobos} (filled triangles)
for the  6\% most central Au+Au are 
compared to $pp$ and $p\bar{p}$ data (open symbols) 
and the NA49 $Pb+Pb$(central 5\%) data 
HIJING1.35 (solid)
 and EKRT (dot-dashed) predictions are also
shown. 
HIJING upper (lower) solid curves are without (with) jet quenching.
The right panel shows the characteristic difference between
the predicted dependence
of the charged multiplicity on the number of participant baryons
for HIJING (solid) and EKRT (dot-dashed).}

\label{fig:phobos}
\end{figure}

The first published data\cite{phobos}
 from RHIC are shown in Fig.\ref{fig:phobos}.
Amazingly, both HIJING and EKRT  predictions are seen to be consistent with
the central multiplicity density data in Au+Au  at $\sqrt{s}=65, 130$ AGeV
as measured by PHOBOS\cite{phobos}. 
However, as emphasized in \cite{wg00}
the scaling as a function of centrality
will soon be able to differentiate between these two very different
pQCD based estimates 
of RHIC initial conditions. In addition, the transverse energy
systematics will provide an independent critical test
of the hypothetical saturation picture\cite{dumimg} as we discuss below.
While it is obviously premature to draw any physics conclusion from these first
data, the consistency of the
increased activity per baryon predicted due to the
onset of mini-jet activity beyond $\sqrt{s}> 100$ AGeV is reasuring.
A critical test of the pQCD framework used above will be the upcoming
RHIC data at $\sqrt{s}=200$ AGeV. HIJING predicts sensitivity to jet quenching
at that energy.

\section{Barometric Measures  of  Collective Dynamics}

The simplest global barometer of collectivity in nuclear reactions is
the A and energy dependence of the transverse energy and charged
particle rapidity density. At RHIC energies, HIJING predicts that
initial transverse energy density in central $A+A$ collisions scales
nonlinearly with A \beq \frac{dE_\perp}{dy} \approx 1 \;{\rm GeV}\;
A^{1.3}\;(1+ \log\frac{\sqrt{s}}{200}) \eeq{et} This leads to about
$0.6$ TeV per unit rapidity at the present $\sqrt{s}=130$ AGeV energy.
In EKRT\cite{Eskola:2000fc}, on the other hand, the gluon saturation
hypothesis predicts an approximate linear A dependence of $E_\perp\sim
A^{1.04}$, and a value several times that of HIJING.  In
Fig.(\ref{fig:mini}), the initial gluon density however grows less
than linear $A^{0.92}$ in that model.  The $A^{1.3}$ scaling of HIJING
with the fixed mini-jet scale $p_0=2$ GeV scale.  This results from
the combined increase of the number of binary interactions as well as
the increased fraction of the energy that originated from the
mini-jets.  The initial energy density in HIJING varies approximately
as 
\beq \epsilon_0\approx 0.6\; {\rm GeV/fm}^3\; A^{0.63}\;(1+
\log\frac{\sqrt{s}}{200}) \; \; .\eeq{ep0} EKRT \cite{Eskola:2000fc}
on the other hand predict $\epsilon(\tau=1/p_{sat})\approx 0.1 A^{0.5}
s^{0.38}$.

If local equilibrium is achieved and maintained, 
then  hydrodynamics predicts 
that   longitudinal boost invariant  expansion
together with $pdV$ work done by the plasma pushing matter down the beam pipe
will cool the plasma and convert some its random transverse
energy into collective longitudinal kinetic energy.
 For an equation of state, $p=c_s^2 \epsilon$,
 this cooling and expansion causes the energy
 density to decrease with proper time as
\beq
\epsilon(\tau)=\epsilon(\tau_0)\left(\frac{\tau_0}{\tau}\right)^{1+c_s^2}
\eeq{bj2}
Entropy conservation leads to a conservation
of $\tau n(\tau)$, where $n$ is the proper parton density. 
At least if the matter is initially deep in the plasma phase, then
(as seen in Fig.3) longitudinal  will be done with $c_s^2\approx 1/3$.
Consequently, the transverse energy per particle
should decrease by a factor 2-3 before freeze-out as\cite{Nayak:2000js}
\beq
e_\perp(\tau)=
\frac{dE_\perp}{dN}=e_\perp(\tau_0)\left(\frac{\tau_0}{\tau}\right)^{c_s^2}
\eeq{work} 
However,  dissipative effects due to rapid expansion and finite
mean free paths reduce considerably the effective pressure\cite{dumimg}.
For the Bjorken expansion, the relaxation time,
$\tau_c=1/(\sigma_{T}n)\propto \tau$, 
then increases with time as $n$ decreases.
Numerical solution
of 3+1 D transport equations with pQCD cross sections,
$\sigma_T\sim 2$ mb, indicate that
dissipation reduces the transverse energy loss
for HIJING initial conditions rather significantly
(see detailed comparisons in \cite{Gyulassy:1997zn,molnar}).

One of the important  experimental 
observations  at the 
SPS  is that at
$\sqrt{s}\sim 20$ AGeV,  $dE_\perp/dy$ as well as $dN/dy$ scale 
 nearly linearly with  wounded nucleon or participant number
($\sim A^{1.07}$) \cite{Schlagheck:2000aq}.
Simple Glauber wounded nucleon models reproduce very
well the nearly linear correlation between $E_\perp$ 
and the veto calorimeter (spectator) energy 
observed in all experiments at SPS\cite{Kharzeev:1997yx}. 
The physics implications of this scaling
depends on what is {\em assumed} for the $A$ dependence
of the initial conditions. One view\cite{Eskola:2000fc} is that
the initial $e_\perp$ scales in just the right way
that after hydrodynamic expansion
 the final $E_T$ and $e_\perp$ always
scale linearly with $A$.
My view 
is that at the SPS
 the linear dependence
arises from  additive nature of soft beam fragmentation together with 
the {\em absence} of $pdV$ work at early times.
If the QGP transition region is just barely
reached, as I believe,
then the softness of the QCD equation of state with
 $c_s^2\ll 1/3$ seen in Fig.(\ref{fig:eos}) and dissipation
conspire to prevent the dense matter from
performing longitudinal work.
However, it is impossible to tell from the data 
whether this observed null effect  is   due
to a low pressure  Hagedorn resonance gas of hadrons or 
to a low pressure lazy 
``plasma'' with $c_s\ll 1$.
At RHIC energies the plasma starts deep in the $c_s\approx 1/sqrt{3}$
Yukawa regime, and 
longitudinal work should become observable in the $E_T$ systematics.

\section{Transverse Collective Flow}

\begin{figure}
\vspace{0in}
\centerline{\epsfxsize= 3 in \epsfysize= 3 in
\epsfbox{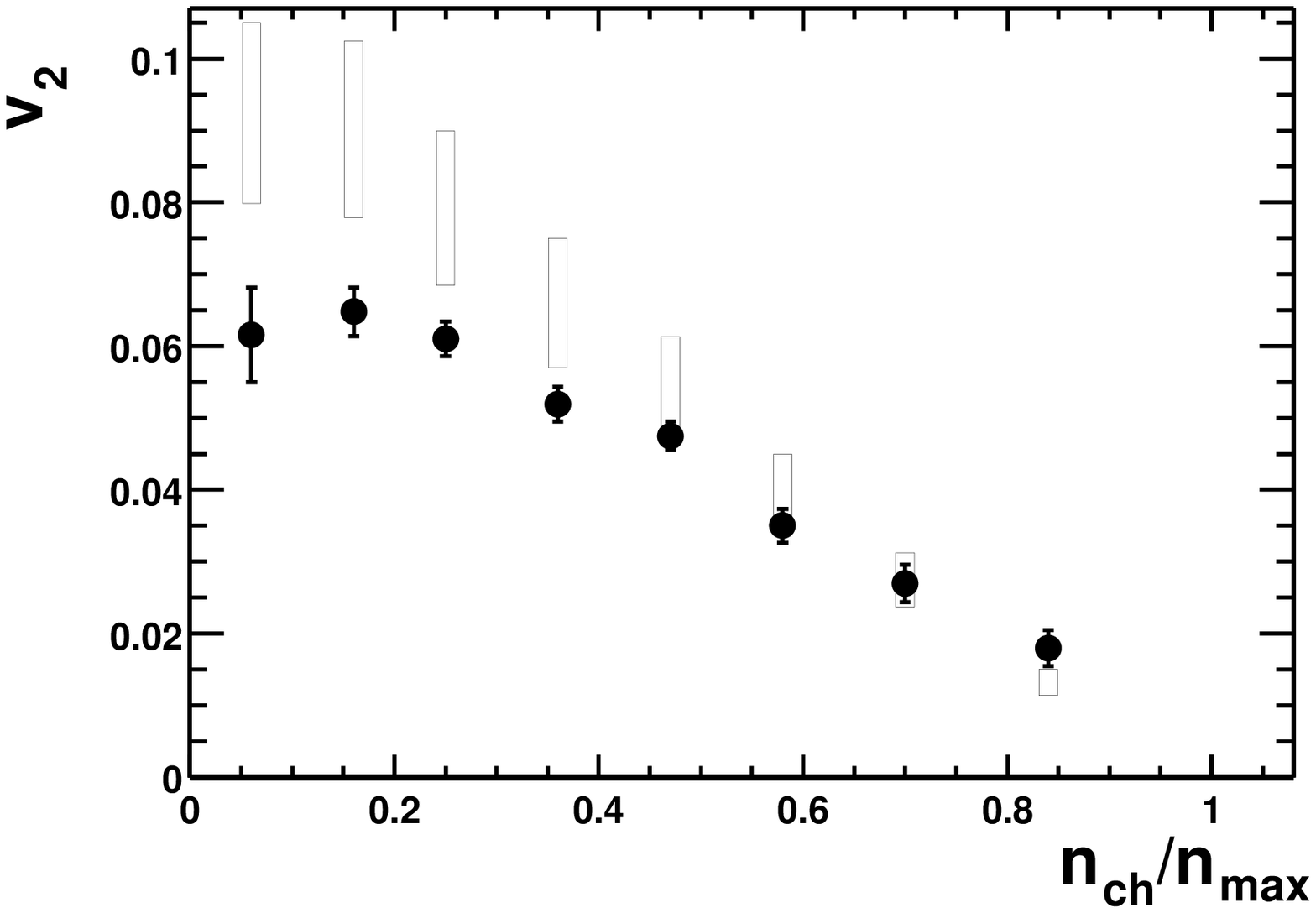}
\hspace{-0.2in}{\hbox{\psfig{figure=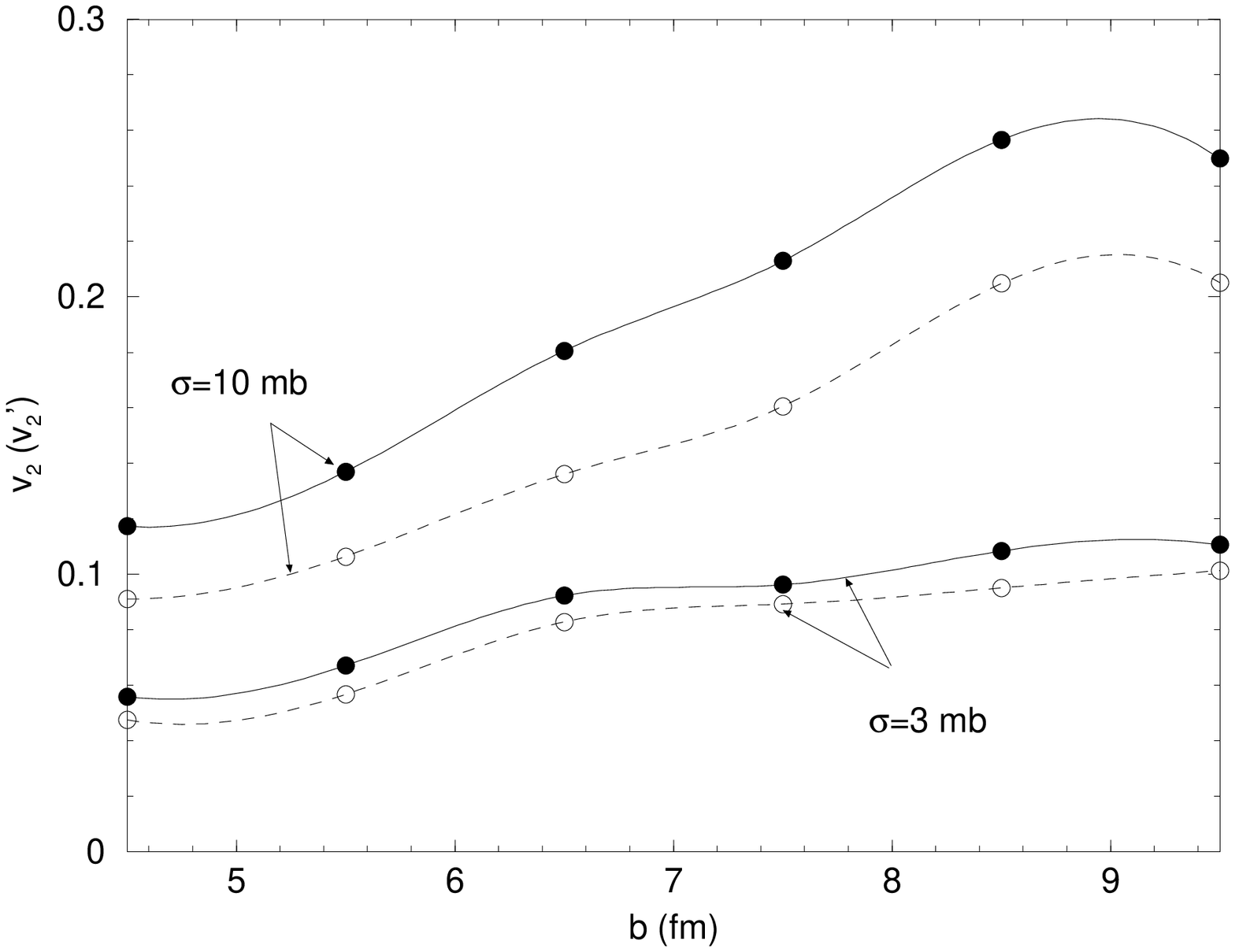,angle=-90,width=9cm}}}}
\caption{(left) First STAR data \protect{\cite{starv2}} 
on the centrality dependence of the azimuthal flow moment $v_2$ 
for $Au+Au$ at 130 AGeV. Estimates of the corresponding spatial asymmetry
are indicated by boxes.
(Right) Predicted dependence of $v_2$ at mid rapidity $Au+Au$ at
RHIC from parton cascade calculations 
ZPC\protect{\cite{Zhang:1999rs}} with HIJING initial conditions
for $\sigma_{gg}=3$ and $10$ mb. (Note $v_2^\prime$ is the $p_\perp$ weighed
form of $v_2$)}
\label{fig:wa98flow}
\end{figure}

An entirely
different measure of barometric collectivity is afforded by the study
of the triple differential distributions, $dN/dyd^2\vp_\perp$. Already
at sub-luminal Bevalac energies ($<1$ AGeV), azimuthally asymmetric
 collective directed and elliptic flow were
discovered long ago. For non central collisions, $b\ne0$, the asymmetric
transverse coordinate profile of the reaction region 
 leads to different gradients of the pressure as a function of
the azimuthal angle relative to beam axis.
  This leads to a ``bounce'' off of projectile and
target fragments in the reaction plane and to azimuthally asymmetric
transverse momentum dependence of particles with short mean free at
mid rapidities.  This phenomenon has now been observed at both
AGS and SPS energies as well\cite{wa98flow}.  It has been predicted
to be there also at RHIC and
LHC energies\cite{Voloshin:2000gs}. 
In Fig.(\ref{fig:wa98flow}) the first STAR data\cite{starv2} on
the centrality dependence of the second  Fourier component, $v_2$, of the
azimuthal flow patterns  is shown:
\beq
\frac{dN}{dyd^2\vp_\perp}\propto 1+ 2v_1 \cos( \phi-\phi_R)+
2v_2 \cos( 2(\phi-\phi_R)) +\cdots \;\; .
\eeq{v2}
These data show that the azymuthal asymmtery is about twice as large at RHIC than at the SPS. Furthermore, the differential azymuthal flow, $v_2(p_\perp)$,
rises linearly with $p_\perp$ up to 2 GeV/c.

The important question is how this type of barometer could serve to
help search for evidence of the QCD transition.  In \cite{jolli1} the
idea was proposed that one could use $v_2$ systematics to search for
the predicted softening of the QCD equation of state.  Hydrodynamics
calculations lead to a factor of two smaller $v_2$ for an equation of
state with a soft critical point as in Fig 3 versus one in which the
speed of sound remains 1$/\sqrt{3}$.  Searches for anomalous $v_2$
dependence as well as $v_1$ are underway\cite{wreis1}. As with the
global barometer, dissipation can of course also simulate a soft
equation of state.  In ref\cite{Zhang:1999rs} we studied the
dependence of $v_2$ on the transport parton cross section for RHIC
conditions.  The results are rather sensitive to the cross sections
and initial densities as shown in Fig(\ref{fig:wa98flow}).  For HIJING
initial conditions, dissipation leads to a significant reduction of
$v_2$ relative to hydrodynamics.

\section{The QGP Stall and Interferometry}

Hadron interferometry has been developed over the last several decades
in heavy ion collisions into a precise tool
to image the space-time region of
the decoupling 4-volume. In \cite{pratt} it was proposed that
a possible  signature of the QGP phase transition
would be a time delay associated with very slow
hadronization. The plasma "burns" into hadronic ashes along
deflagration front that moves very slowly if the entropy drop across the
transition is large\cite{vanH}.
In \cite{Rischke:1996em} we calculated
the 2+1D evolution of a Bjorken
cylinder with time and transverse coordinate using
hydrodynamics  with different equations of state.
For an initial energy density $\sim 20$ GeV/fm$^3$
a time delay up to a factor of two was found
even for a continuous transition as long as the entropy jump was
$\sim 10$. This QGP "stall" of the transverse expansion
is due to the small  speed of sound in the mixed phase.
It should be readily observable in pion interferometry
by comparing $R_{out}$ and $R_{side}$ radii.

The time delay is a robust generic
signature of a rapid cross over transition  of the entropy density.
In particular the "stall" is expected  even for a smooth
cross-over transition 
as long as the width $\Delta T/T_c <0.1$. However, its magnitude also
depends strongly on the entropy drop across that region. 
 Unfortunately, as noted before
lattice QCD has not yet been able to resolved the hadronic world below $T_c$
due to technical problems. If the entropy jump in nature is less than a factor
of three,
then the stall signature would be much more difficult
to  observe.

High statistics measurements of pion and 
kaon interferometry searches for time
delay at AGS and SPS have come up empty handed thus far.  No time delay 
has ever
been observed in any nuclear reactions thus far. 
This could be due to (a) the absence of
a large rapid entropy drop in real QCD, or (b)
to unfavorable kinematic conditions at AGS and SPS energies.  From the
hydrodynamic calculations in \cite{Rischke:1996em},
 it was found that a large 
time delay signal requires a favorable initial condition
with initial temperature $\sim 2 T_c$. For too high an initial temperature
(as at LHC) the large transverse collective flow that develops
prevents a stall from happening. For too low a temperature (as at SPS) the time delay is suppressed because the system spends too
little time in the mixed phase as a result of longitudinal expansion.
Therefore, RHIC is the most likely energy regime where
a QGP stall may be observable. Recent  work
\cite{Bernard:1997bq} has also shown that high $p_\perp$ kaon interferometry 
is an  especially sensitivity probe of
time delay. If  observed, a time delay signature
would be one of the most powerful indicators
of novel collective
behavior that can only arise if the produced matter has an  anomalous
(softest point) equation of state.

\section{J/Psi Suppression}

In 1986, Matsui and Satz proposed an intriguing direct measure 
of the transmutation of the $q\bar{q}$ forces shown in Fig. 1.
The idea was that $J/\psi$ can form in the vacuum because
the confining Luscher potential can bind a $c\bar{c}$ pair into
that vector meson. If that pair were placed in a hot medium in which
the 
potential is  screened, 
then above the temperature where the screening length is
smaller than the $J/\psi$ radius, the $c\bar{c}$ would 
become unbound. The charm quarks would then emerge from the reaction region
as an open charm $D\bar{D}$ pair. 

$J/\psi$ suppression was soon  seen  in 1987 in $O+U$
 reactions by NA38. Since then this smoking gun has
(unfortunately) never stopped smoking! $J/\psi$ suppression seems to
be as ubiquitous
as the Yukawa potential. It was subsequently
 observed in $p+A$ reactions as well.
 High mass Drell-Yan pairs, 
on the other hand, formed via
 $q\bar{q}\rightarrow \ell \bar{\ell}$
was observed to
scale  perfectly linearly with the number of binary collisions.
This is because 
lepton pairs suffer no final state interactions
and the quark initial state (Cronin) interactions are invisible
in $p_\perp$ integrated DY yields.

The ``normal, conventional'' $J/\psi$   suppression leading to a  $(AB)^{0.9}$ dependence,
is presumed to be due to a mechanism that is independent of QGP production. 
In 
Pb+Pb, NA50 observed an excess 25\% suppression of $J/\psi$ above that
normal expectation.
 NA50  therefore claimed that this enhanced suppression
is finally the sought after smoking gun of QGP formation\cite{Abreu:2000ni}.
\begin{figure}
%
\centerline{\epsfxsize= 3.3 in \epsfbox{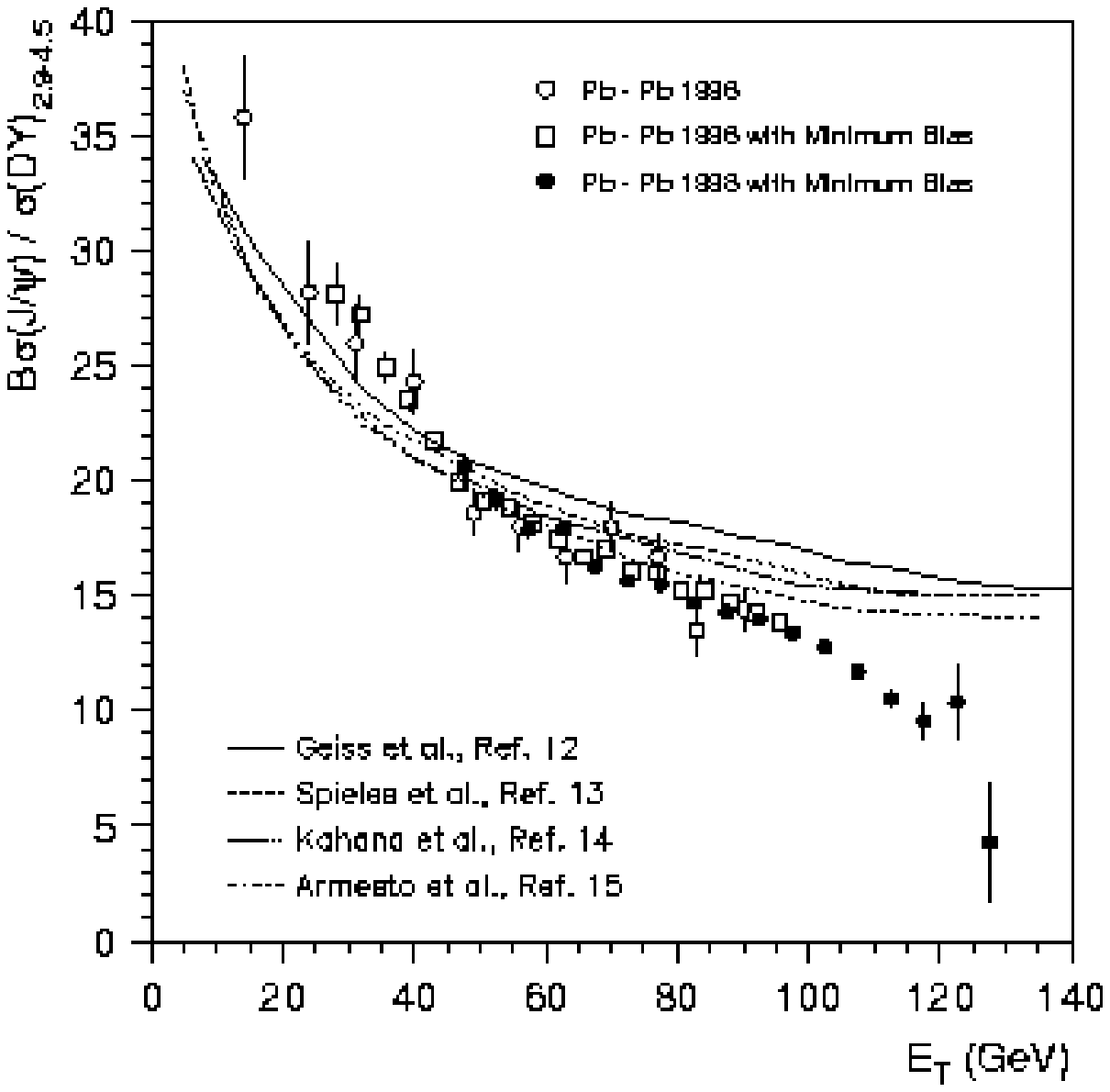}
\epsfxsize= 3.3 in 
  \epsfbox{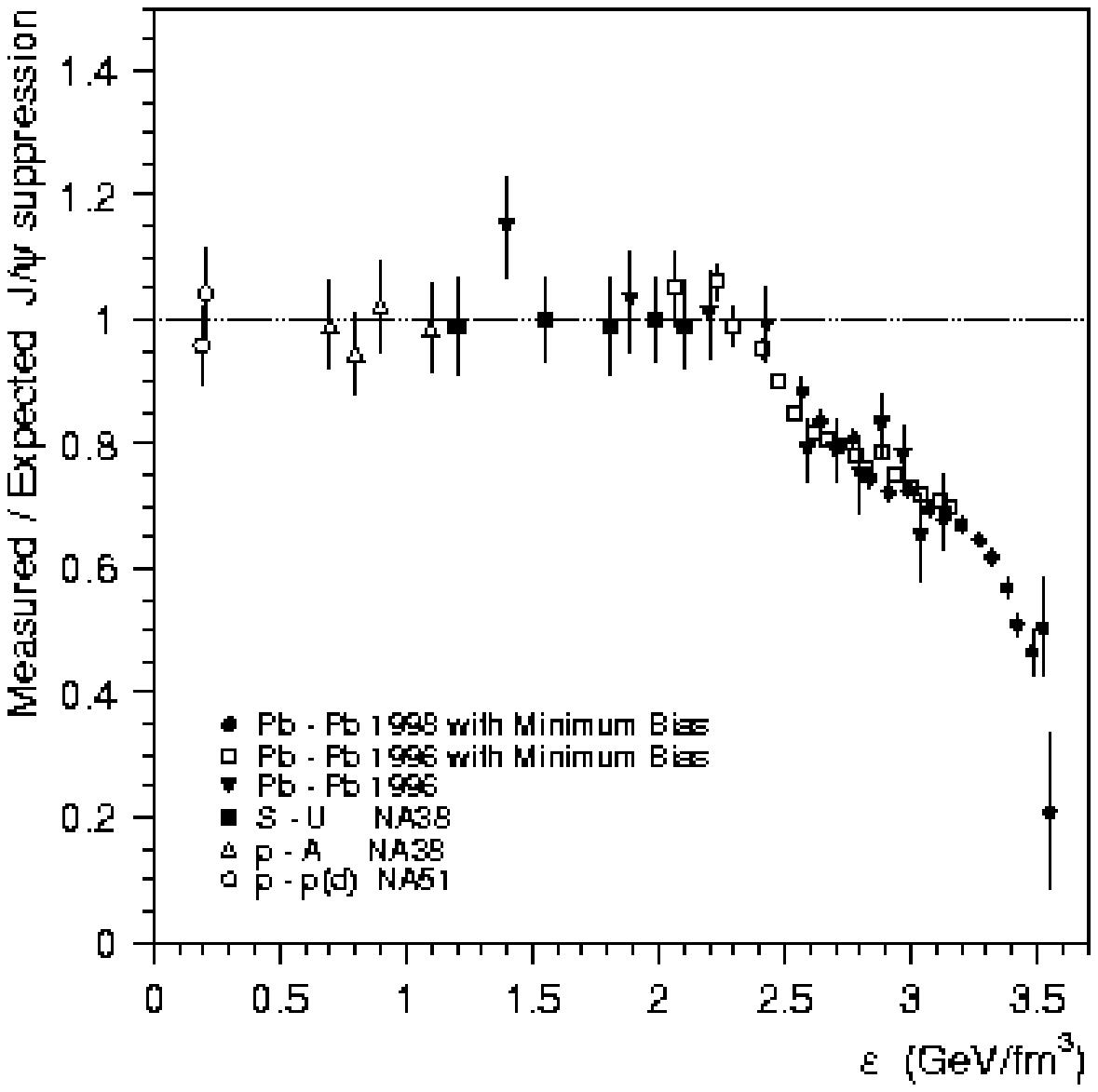}
}
\caption{``Evidence for deconfinement of quarks and gluons'' 
claimed by NA50 in Ref.\protect{\cite{Abreu:2000ni}}.
The curves on left show transport theory estimates for hadronic final state 
dissociation. The enhanced suppression relative to $\exp(-\sigma_N \rho_0 L)$
nuclear suppression observed in $p+A$ and $S+U$ is shown versus
a very rough estimate of the unobserved initial energy density.}
\label{fig:psina50}
\end{figure} 

While the deviation from the  empirical $(AB)^{0.9}$
scaling is convincing, the dynamical origin of the effect is
far from settled  in my opinion. As shown by the several curves in 
Fig.(\ref{fig:psina50}),
a $\sim$50\% drop in the $J/\psi$ yield as a function of $E_T$
is consistent
with final state co-moving hadronic absorption. While the detailed wiggles
are not reproduced, large theoretical
uncertainties  about several key dynamical 
 ingredients preclude more precise comparisons at this time. 
It is important to emphasize that  plasma scenario models 
suffer at least as large theoretical uncertainties as
the so called ``conventional hadronic''
models. 
Key uncertain elements include (1) the proper QCD  treatment of
cold nuclear absorption responsible for the nonplasma ${AB}^{0.9}$ suppression
observed in $p+A$, (2) the unknown hadronic $M+\psi\rightarrow D\bar{D} X$
reaction rates, and (3) the detailed density evolution, $\epsilon(\vx,t)$.

\begin{figure}
\vspace{-2in}
\centerline{\epsfxsize= 5in \epsfbox{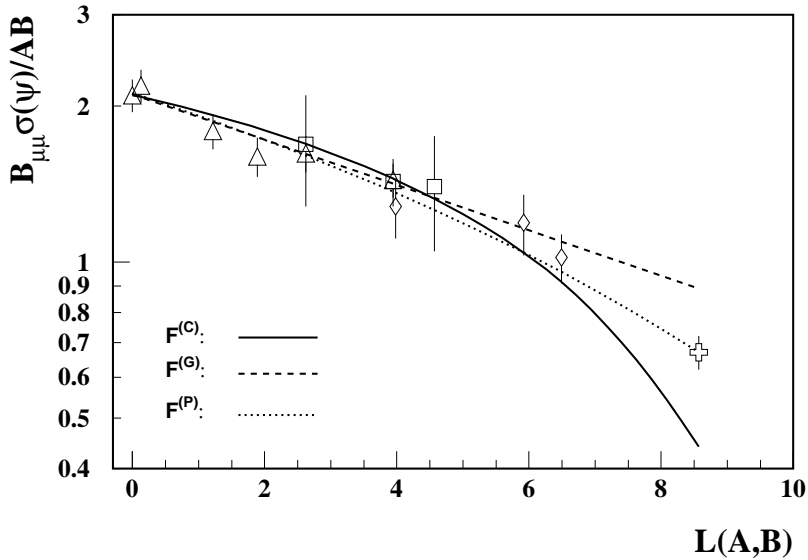}
}
\vspace{-2in}
\caption{Sensitivity of ``pre-resonance" $J/\psi$ nuclear
absorption to details  of color octet models of $c\bar{c}$
interactions in a nuclear medium from Qiu et al\protect{\cite{Qiu:1998rz}}.
The curve marked P for power law accounts for the anomalous absorption seen in
$Pb+Pb$ and deviated from Gaussian Glauber-like G expectations.}
\label{fig:qui}
\end{figure}

For example, only a schematic  ``octet model" of pre-hadronization
$c\bar{c}$ exists at present 
to calculate the so called ``normal'' nuclear suppression.
That this is model may not be suffiencient to
adequately account for that component can be seen in the work of  
Ref.\cite{Qiu:1998rz}. The observed $J/\psi$ production cross section
(ignoring  final state interactions) may be expressed as
\beq
\sigma_{AB\rightarrow\psi X}=\int d\sigma_{AB\rightarrow c\bar{c}}
F_{c\bar{c}\rightarrow\psi}(q^2)
\eeq{qui}
where $F$ is the formation probability of the $\psi$ from a $c\bar{c}$
that emerges from the cold nuclear target with an invariant mass
$q^2<4M_D^2$. If the pre-resonance $c\bar{c}$ 
pair multiply scatters in the nucleus,
then $q^2\rightarrow q^2 + \delta q^2 (\sigma\rho L)$
increases linearly with nuclear thickness as shown by the  curve G in
 Fig.(\ref{fig:qui}). This leads to an approximate
exponential suppression that can account for the approximate  Glauber
nuclear absorption factor ansatz, $\exp(-\sigma_{eff}\rho L)$.
This Gaussian model  can thus 
account for the observed $(AB)^{0.9}$ scaling for light projectiles. 
{However},
it was shown in \cite{Qiu:1998rz} that if the power law tails due to induced
radiation in the medium are included (resulting
from the multiple 
Rutherford rescattering of the color octet $c\bar{c}$)
, then an additional nonlinear suppression
in the nuclear thickness $L$ {\em could} result (curve P). This is  because
induced gluon radiation provides a way to increase the invariant mass of the
pre-resonance 
$c\bar{c}$ that can further reduce the probability 
for the pair to fit inside the $\psi$ wavefunction.
While this model is also rather schematic, 
I feel that it captures one of the possibly important
elements of pre-formation {\em transient} dynamical effects in nuclei
that may account for anomalous nonlinear suppression
in the thickest nuclei.

The claim of  anomalous
suppression cannot  therefore rest  merely 
on generic enhanced suppression in $Pb$.
It rests  on the possible existence  of singular
``step-like" structure of the suppression pattern.
The evidence for "steps" is however the weakest link experimentally
because a rigorous $\chi^2$ test including the substantial systematic
errors in the $E_T$ scale has yet to be performed.
Nevertheless,
if the ``step-like" suppression pattern survives further experimental
scrutiny, it would certainly be the most dramatic nonlinearity observed at SPS.
Experimentally, the claims would carry considerable more weight if
similar "step-wise" patterns were observed in other systems ,e.g.
$Xe+Xe$ suitably shifted in $E_T$ due to the expected smaller energy
densities achieved. I would put such an experiment highest on the SPS
priority list. At RHIC, the striking prediction by H. Satz at
QM99\cite{Bass:1999zq}, was that under RHIC conditions the higher
energy density should lead to the same step-wise pattern 
in the lighter nuclear $Cu+Cu$ interactions. PHENIX will provide a
definitive test of this prediction in a few years.

\begin{figure}
%
\centerline{\epsfxsize= 3.2 in \epsfbox{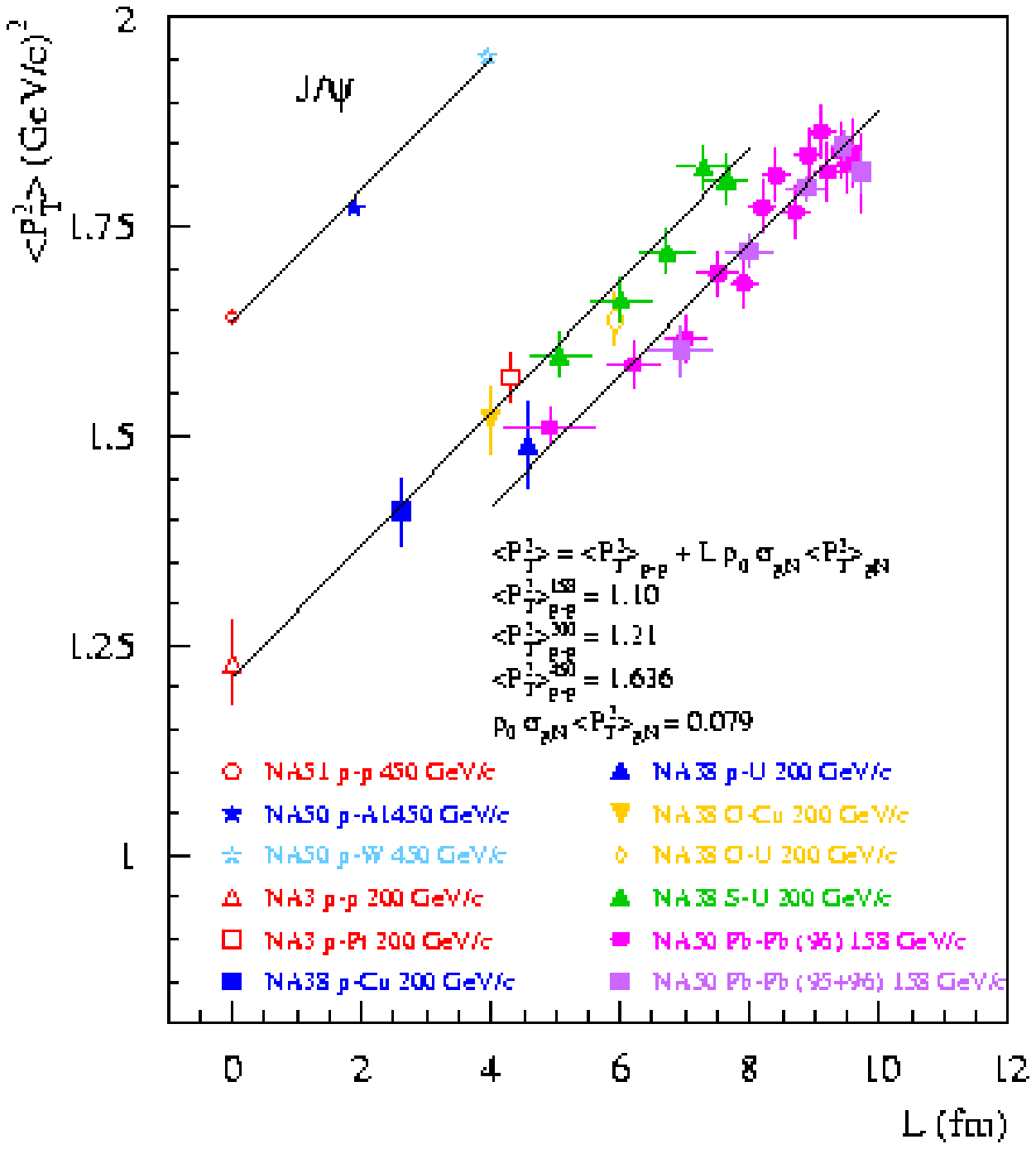}
\epsfxsize= 3.3 in 
  \epsfbox{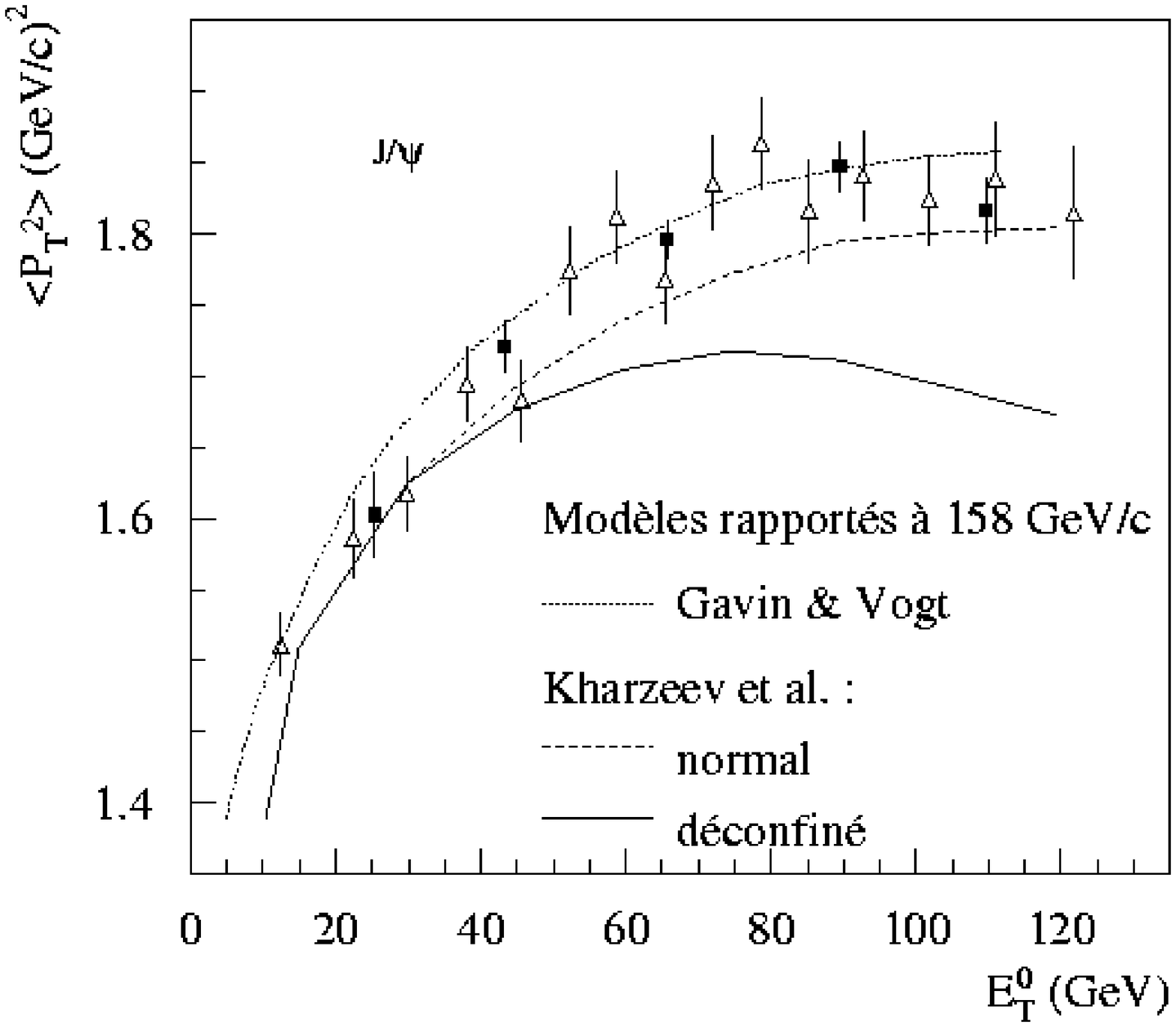}
}
\caption{The mean transverse momentum of the $J/\psi$ 
\protect{\cite{Drapier98}}
shows clear evidence of multiple scattering in the nuclear
target as expected\protect{\cite{Gavin:1988tw}} from  Glauber theory. 
Even in $Pb+Pb$ the increase of the $p_T$ is understood from the same
mechanism in contradiction\protect{\cite{Nagle:1999ms}} 
to predictions based on the plasma
scenario, where at high $E_T$,  the surviving $\psi$ are expected to have been
produced only near the nuclear surface 
regions\protect{\cite{Kharzeev:1997ry}}.
}
\label{fig:psipt}
\end{figure}

One  significant inconsistency  
with the present plasma scenario interpretation
is its failure to account for  the observed $E_T$ dependence of the $J/\psi$ $p_\perp$
spectra in  Fig.(\ref{fig:psipt}).
 Standard Glauber multiple collisions lead to a random walk
in transverse momentum that are expected enhance the $J/\psi$
transverse momentum as \cite{Gavin:1988tw}
\beq
\langle p_\perp^2\rangle_{AB}= \langle p_\perp^2\rangle_{pp} + 
\frac{L}{\lambda} \delta p_\perp^2
\;\; .\eeq{ptpsi}
This is found to hold\cite{Drapier98} experimentally
in all reaction {\em including} $Pb+Pb$.
In contrast, in the plasma scenario\cite{Kharzeev:1997ry}, 
only those $\psi$ are
expected  to be observed which are produced 
near the surface, where the effective nuclear depth
$L$ is small. Thus, the prediction as shown in Fig(\ref{fig:psipt}) was
that the $\langle p_\perp^2\rangle$ should begin to DECREASE with increasing
$E_T$. This was not observed\cite{Drapier98,Nagle:1999ms}.

\section{The High $p_\perp$ Frontier}

One of the most exiting new physics areas that 
will become accessible at RHIC 
is the study of high transverse momentum (short
wavelength) jet probes. The rates of jet production and its fragmentation
in the vacuum are well understood. The new jet physics at RHIC will
be the study of partonic interactions at extreme densities
through the phenomenon of jet quenching\cite{gptw}.
Final state interactions of a jet in a dense QGP are
expected to induce a large radiative energy loss\cite{mgxw}.
In fact, BDMPS\cite{bdms} discovered
 that non-Abelian energy loss
is in fact non-linear as a function of the thickness of the medium.
Tests of this and other aspects of 
 non-Abelian multiple  collision dynamical phenomena will soon be possible 
at RHIC\cite{glv1b}.

\begin{figure}
%
\centerline{
{\hbox{\psfig{figure=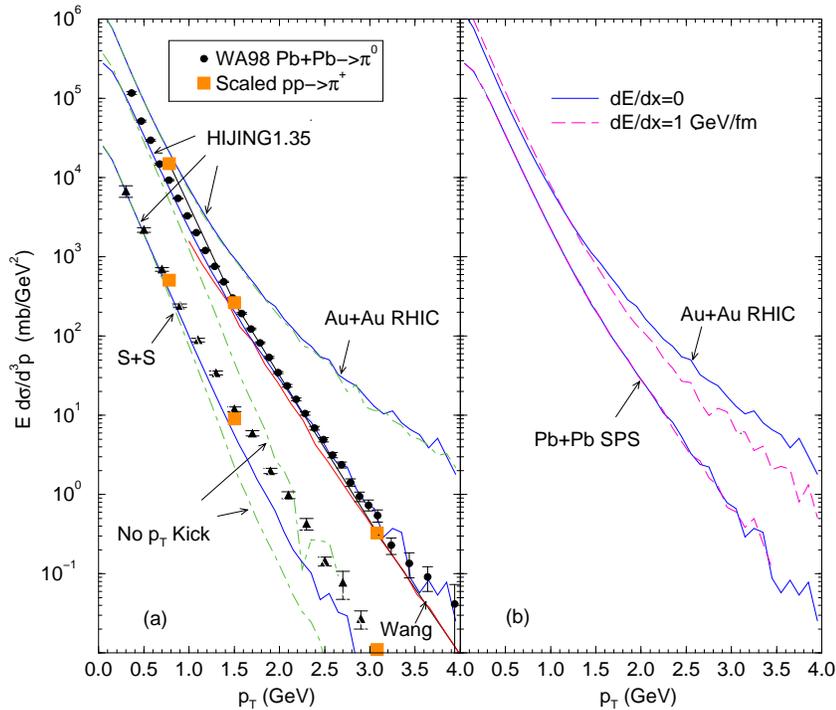,angle=-90,width=11cm}}}
}
\caption{Jet quenching at SPS 
vs RHIC from ref.\protect{\cite{Gyulassy:1998nc}} compared to WA98 data.
At RHIC the power law tail extends finally far enough above 
 the nonperturbative
``noise" to make jet quenching  observable (long dashed in right panel).
At SPS the steep high $p_\perp$ tails are too sensitive multiple soft
scatterings.}
\label{fig:jets}
\end{figure}

At lower SPS energies, this  physics is out of reach because
nonperturbative multiparticle production physics (e.g., 
soft multiple $p_\perp$ kicks) dominates
as shown in Fig.(\ref{fig:jets})
from ref. \cite{Gyulassy:1998nc}. HIJING happens to fit the WA98
data with or without jet quenching at SPS energies. 
While at the SPS no clean separation of
soft and hard dynamics is kinematically possible,
at RHIC energies, the high $p_\perp$ power law tails of the single inclusive
distributions stick out far enough above the  soft physics "noise" that
sensitivity to the form of the non-Abelian $dE/dx$ is expected 
as shown in Fig.(\ref{fig:jets}).
This problem is closely related also to the problem
of pre-resonance $J/\psi$ absorption discussed previously\cite{Qiu:1998rz}.
Understanding jet quenching is also necessary to develop more accurate
covariant parton transport theories\cite{Gyulassy:1997zn,molnar}.

\section{Summary}

The next Yukawa phase of QCD is awaiting discovery. The SPS data have
provided many intriguing indirect hints that new physics is already
operating near the transition region.  Many claims and counter claims
remain at SPS energies because both hadronic and partonic models have
partial overlapping domains of validity.  This ``duality" is analogous
to the problem of interpreting the R factor in $e^+e^-$ collisions
below $\surd s<10$ GeV. The ratio of hadronic to leptonic production
cross sections only reaches the magic 11/3 of pQCD above the threshold
regions for $s\bar{s},c\bar{c},b\bar{b}$.  Near the threshold regions
conventional vector mesons models of nonperturbative hadronic physics
provide an adequate ``dual'' description of the physics.  The simpler
continuum partonic description is only applicable far above that
region.  Similarly, reactions at SPS energies are near the threshold
region where resonances become too broad due to multiple interactions.
In nuclear reactions, the continuum parton QGP description can only be
expected to apply far above that region as produced at collider
energies.

At RHIC with a factor of ten increase in the initial energy density,
only a continuum parton description of the initial conditions is
tenable. The dense partonic matter so formed also will have much more
time to evolve and produce collective signatures of its existence.
Furthermore, a factor of ten smaller wavelength (jet) probes will
finally allow experimentalists to resolve (i.e. see) individual quark
and gluon degrees of freedom of that plasma. Direct observation of
longitudinal work, transverse azimuthal collectivity, time delay,
step-wise $J/\psi$ suppression in $Cu+Cu$, and jet quenching should,
among other signatures, allow experimentalists to measure the
properties of the partonic Yukawa phase at RHIC.

\section*{Acknowledgements}
This work was supported by the Director, Office of Energy Research,
Division of Nuclear Physics of the Office of High Energy and Nuclear Physics
of the U.S. Department of Energy under Contract No. DE-FG-02-93ER-40764.


\begin{thebibliography}{99}
\bibitem{Yuk35}
H. Yukawa, Proc.~Phys.-Math.~Soc.~Jpn.~Ser.3, {\bf 17} (1935) 48.



\bibitem{Bonn} R. Machleidt, {\it The Meson Theory of Nuclear Forces and
Nuclear Structure}, in {\it Advances in Nuclear Physics}, {\bf 19},
eds. J.W. Negele and E. Vogt,(Plenum Press, New York, 1989), chap.2.


\bibitem{Debye}
P. Debye and E. H\"uckel, Z. Phys. {\bf 24} (1923) 185.

\bibitem{Lusch}
M. L\"uscher, K. Symanzig, and P. Weise, NPB 173 (1980) 365.

\bibitem{Gao} M. Gao, Phys. Rev. D40 (1989) 2708.

\bibitem{Kaczmarek:1999mm}
O.~Kaczmarek, F.~Karsch, E.~Laermann and M.~Lutgemeier,
hep-lat/9908010.


\bibitem{Rebhan:1994mx}
A.~K.~Rebhan,
Nucl.\ Phys.\  {\bf B430}, 319 (1994)

\bibitem{Bernard:1997cs}
C.~Bernard {\it et al.}  [MILC Collaboration],
Phys.\ Rev.\  {\bf D55}, 6861 (1997)
[hep-lat/9612025].



\bibitem{McLerran:1994ni}
L.~McLerran and R.~Venugopalan,
Phys.\ Rev.\  {\bf D49}, 2233 (1994)
[hep-ph/9309289].

\bibitem{Eskola:2000fc}
K.~J.~Eskola, K.~Kajantie, P.~V.~Ruuskanen and K.~Tuominen,
Nucl.\ Phys.\  {\bf B570}, 379 (2000)
[hep-ph/9909456].

\bibitem{Eskola:1989yh}
K.~J.~Eskola, K.~Kajantie and J.~Lindford,
Nucl.\ Phys.\  {\bf B323}, 37 (1989).

\bibitem{Blaizot:1987nc}
J.~P.~Blaizot and A.~H.~Mueller,
Nucl.\ Phys.\  {\bf B289}, 847 (1987).
\bibitem{Gyulassy:1997vt}
M.~Gyulassy and L.~McLerran,
Phys.\ Rev.\  {\bf C56}, 2219 (1997)
[nucl-th/9704034].

\bibitem{Wang:1991ht}
X.~Wang and M.~Gyulassy,
Phys.\ Rev.\  {\bf D44}, 3501 (1991).
\bibitem{Wang:1997yf}
X.~Wang,
Phys.\ Rept.\  {\bf 280}, 287 (1997)
[hep-ph/9605214].


\bibitem{cernhype} CERN Press Release 2/8/00:
\verb+http://press.web.cern.ch/Press/Releases00/+\\
\verb+PR01.00EQuarkGluonMatter.html+,\\
\verb+http://cern.web.cern.ch/CERN/Announcements/2000/NewStateMatter/+,\\
and U. Heinz these proceedings.
\bibitem{Gyulassy:1998nc}
M.~Gyulassy and P.~Levai,
Phys.\ Lett.\  {\bf B442}, 1 (1998)
[hep-ph/9807247].


\bibitem{Bass:1999vz}
S.~A.~Bass, M.~Gyulassy, H.~Stocker and W.~Greiner,
J.\ Phys.\ G {\bf G25}, R1 (1999)
[hep-ph/9810281].


\bibitem{phobos}
B.B.~Back et al., PHOBOS Collaboration, hep-ex/0007036; 
\bibitem{wg00}
X.~Wang and M.~Gyulassy, nucl-th/0008014.

\bibitem{Nayak:2000js}
G.~C.~Nayak, A.~Dumitru, L.~McLerran and W.~Greiner,
hep-ph/0001202.

\bibitem{dumimg}
A.~Dumitru and M.~Gyulassy,
hep-ph/0006257.


\bibitem{Gyulassy:1997zn}
M.~Gyulassy, Y.~Pang and B.~Zhang,
Prog.\ Theor.\ Phys.\ Suppl.\  {\bf 129} (1997) 21;
Nucl.\ Phys.\  {\bf A626}, 999 (1997)
[nucl-th/9709025].
\bibitem{molnar} D. Molnar, M. Gyulassy, e-Print Archive: nucl-th/0005051.

\bibitem{Schlagheck:2000aq}
H.~Schlagheck  [WA98 Collaboration],
Nucl.\ Phys.\  {\bf A663\&664}, 725 (2000)
[nucl-ex/9909005].

\bibitem{Kharzeev:1997yx}
D.~Kharzeev, C.~Lourenco, M.~Nardi and H.~Satz,
Z.\ Phys.\  {\bf C74}, 307 (1997)
[hep-ph/9612217].


\bibitem{wa98flow} M. Aggarwal et al WA98, Nucl. Phys. A610 (1996) 200c.

\bibitem{Voloshin:2000gs}
S.~A.~Voloshin and A.~M.~Poskanzer,
Phys.\ Lett.\  {\bf B474}, 27 (2000)

\bibitem{starv2}
K.~H.~Ackermann {\it et al.}  [STAR Collaboration],
nucl-ex/0009011.










\bibitem{jolli1}
J.-Y. Ollitrault, Phys. Rev. D {\bf 46} (1992) 229.
\bibitem{wreis1}
J.-Y. Ollitrault, Nucl. Phys. {\bf A590} (1995) 561c;
W. Reisdorf and H.G. Ritter, Annu. Rev. Nucl. Part. Sci. {\bf 47} (1997) 663.

\bibitem{Zhang:1999rs}
B.~Zhang, M.~Gyulassy and C.~M.~Ko,
Phys.\ Lett.\  {\bf B455}, 45 (1999)
[nucl-th/9902016].

\bibitem{pratt}
S.\ Pratt, Phys.\ Rev.\ C 49 (1994) 2722, Phys.\ Rev.\ 
D 33 (1986) 1314.
\bibitem{vanH} 
L.\ Van Hove, Z.\ Phys.\ C 21 (1983) 93, \\
M.\ Gyulassy, K.\ Kajantie, H.\ Kurki--Suonio, L.\ McLerran, 
Nucl.\ Phys.\ B 237 (1984) 477.


\bibitem{Rischke:1996em}
D.~H.~Rischke and M.~Gyulassy,
Nucl.\ Phys.\  {\bf A608}, 479 (1996)
[nucl-th/9606039].

\bibitem{Bernard:1997bq}
S.~Bernard, D.~H.~Rischke, J.~A.~Maruhn and W.~Greiner,
Nucl.\ Phys.\  {\bf A625}, 473 (1997)
[nucl-th/9703017].

\bibitem{Abreu:2000ni}
M.~C.~Abreu {\it et al.}  [NA50 Collaboration],
Phys.\ Lett.\  {\bf B477}, 28 (2000).


\bibitem{Qiu:1998rz}
J.~Qiu, J.~P.~Vary and X.~Zhang,
hep-ph/9809442.

\bibitem{Drapier98} O. Drapier, Thesis,  Université Lyon-I, 1998

\bibitem{Nagle:1999ms}
J.~L.~Nagle and M.~J.~Bennett,
Phys.\ Lett.\  {\bf B465}, 21 (1999)
[nucl-th/9907004].



\bibitem{Gavin:1988tw}
S.~Gavin and M.~Gyulassy,
Phys.\ Lett.\  {\bf B214}, 241 (1988).

\bibitem{Kharzeev:1997ry}
D.~Kharzeev, M.~Nardi and H.~Satz,
Phys.\ Lett.\  {\bf B405}, 14 (1997)
[hep-ph/9702273].


\bibitem{Bass:1999zq}
S.~A.~Bass {\it et al.},
Nucl.\ Phys.\  {\bf A661}, 205 (1999)
[nucl-th/9907090].

\bibitem{gptw}
        M.~Gyulassy, M.~Pl\"umer, M.H. Thoma and X.-N.~Wang,
        Nucl. Phys. A {\bf 538} (1992) 37c;
 X.-N.~Wang and M.~Gyulassy, Phys. Rev. Lett.
        {\bf 68} (1992) 1480.
\bibitem{mgxw}M.~Gyulassy and X.-N.~Wang, Nucl. Phys. B {\bf 420}
        (1994) 583.
\bibitem{bdms} 
         R.~Baier, Yu.L.~Dokshitzer, A.H.~Mueller and
         D.~Schiff,  Nucl. Phys. B {\bf 531} (1998) 403
and refs therein.
\bibitem{glv1b}
         M.~Gyulassy, P.~L\'evai, I.~Vitev,
         Nucl. Phys. B {\bf 571}  (2000) 197; e-Print nucl-th/0005032 ;
nucl-th/0006010.

\end{thebibliography}
\end{document}